\documentclass[english,journal]{IEEEtran}
\usepackage{amssymb}
\usepackage[cmex10]{amsmath}
\usepackage{stmaryrd}
\usepackage[T1]{fontenc}
\usepackage[latin9]{inputenc}
\usepackage{verbatim}
\usepackage{units}
\usepackage{amsthm}
\usepackage{graphicx}
\usepackage{esint}
\usepackage{float,flafter}
\usepackage{array}
\usepackage{cite}

\theoremstyle{plain}
\newtheorem{thm}{\protect\theoremname}
\newtheorem{remark}[thm]{Remark}

\newtheorem{proposition}[thm]{Proposition}

\usepackage{psfrag}
\usepackage{epsfig}
\usepackage{dsfont}

\usepackage{babel}

\usepackage{subfigure}

\providecommand{\theoremname}{Theorem}

\newcounter{MYtempeqncnt}

\usepackage{tikz,pgf,pgfplots}
\usepgflibrary{shapes}
\usetikzlibrary{arrows,shapes,chains,matrix,positioning,scopes,petri}

  \newif\ifdraft
  \drafttrue

\providecommand{\tabularnewline}{\\}

\begin{document}
\newcommand{\myScl}{0.735}

\newif\ifoneCol
\oneColfalse

\global\long\def\good{\textup{\texttt{g}}}
\global\long\def\bad{\textup{\texttt{b}}}

\pgfdeclarelayer{background}
\pgfdeclarelayer{foreground}
\pgfsetlayers{background,main,foreground}
\tikzstyle{state1}=[rectangle, draw=black,rounded corners, fill=white, text centered, anchor=north, text=black, text width={1cm},  minimum height={2cm}]
\tikzstyle{state}=[circle, draw=black, fill=white, text centered, anchor=north, text=black]
\tikzstyle{state2}=[circle, draw=black, fill=white, text centered, anchor=north, text=black,dashed]

\ifoneCol
\title{Delay-Sensitive Wireless Data Transmissions: Queueing Behavior and \\Optimal Code Parameters}
\else
\title{\scalebox{0.97}{Delay-Sensitive Communication over Fading Channel:}\\ \scalebox{0.97}{Queueing Behavior and Code Parameter Selection}}
\fi

\author{\IEEEauthorblockN{Fatemeh Hamidi-Sepehr, \emph{Student Member, IEEE},
Henry D. Pfister, \emph{Senior Member, IEEE},\\
Jean-Francois Chamberland, \emph{Senior Member, IEEE}}
\thanks{This material is based upon work supported by the National Science Foundation (NSF) under Grants No.~0830696, No.~0747363 and No.~0747470.
Any opinions, findings, and conclusions or recommendations expressed in this material are those of the author(s) and do not necessarily reflect the views of the National Science Foundation.

The authors are with the Department of Electrical and Computer Engineering, Texas A\&M University, College Station, TX 77843, USA (emails: f\_hamidisepehr@tamu.edu; hpfister@tamu.edu; chmbrlnd@tamu.edu).}
}

\maketitle

\begin{abstract}
This article examines the queueing performance of communication systems that transmit encoded data over unreliable channels.
A fading formulation suitable for wireless environments is considered where errors are caused by a discrete channel with correlated behavior over time.
Random codes and BCH codes are employed as means to study the relationship between code-rate selection and the queueing performance of point-to-point data links.
For carefully selected channel models and arrival processes, a tractable Markov structure composed of queue length and channel state is identified.
This facilitates the analysis of the stationary behavior of the system, leading to evaluation criteria such as bounds on the probability of the queue exceeding a threshold.
Specifically, this article focuses on system models with scalable arrival profiles, which are based on Poisson processes, and finite-state channels with memory.
These assumptions permit the rigorous comparison of system performance for codes with arbitrary block lengths and code rates.
Based on the resulting characterizations, it is possible to select the best code parameters for delay-sensitive applications over various channels.
The methodology introduced herein offers a new perspective on the joint queueing-coding analysis of finite-state channels with memory, and it is supported by numerical simulations.
\end{abstract}

\section{Introduction}
\label{section:Introduction}

Contemporary wireless communication systems must be designed to accommodate the wide range of applications that compose today's digital landscape.
Modern mobile devices should be able to support heterogeneous data flows with a variety of delay and bandwidth requirements.
While point-to-point channels have received much attention in the past, the asymptotic approaches favored by classical information theory offer only limited insights on efficient designs in the context of delay-sensitive communications.
Indeed, real-time traffic and live interactive sessions are typically subject to very stringent delay requirements.
Such constraints can hardly be captured by asymptotic regimes where block lengths and, consequently, delay become unbounded.
At this point, it is important to note that several recent contributions to communication theory seek to address the tradeoffs between average power, throughput and delay \cite{Bettesh-it06,Cao-isit08,Li-wcmc10,Rajan-globecom01,Ahmed-Asilomar04}. 
Still, many such articles make idealized assumptions about the performance of coded transmissions.
These assumptions are often reasonable for long codewords, but they are not necessarily justified for low-latency communication over channels with memory.
In this work, we study the impact of certain coding strategies on the queueing performance of finite-state channels with and without memory.
This is accomplished without resorting to characteristic, simplifying assumptions about the operation of coded transmissions.

Before diving into the details of the systems we wish to analyze, we present a brief survey of pertinent prior research contributions.
Forward error-correcting codes have historically played an instrumental role in digital communication systems by providing protection against channel uncertainties.
For instance, it is well-known that for rates below capacity, one can improve transmission reliability by increasing the block length of a code.
There is a tradeoff between the improvements offered by low-rate codes and the payload reduction associated with an increase in redundancy.
Finding a suitable balance between these two intertwined considerations is a fundamental pursuit in coding theory.
The Shannon capacity, for instance, characterizes the maximum achievable throughput a channel can support subject to an asymptotic reliability constraint as block length tends to infinity~\cite{Gallager-1968}.

Due to the delay requirements of certain modern applications, one may be forced to employ schemes with short codewords.
While sometimes necessary, short codes can preclude the concentration of empirical measures for errors and channel state occupancy.
This may, in turn, produce excess decoding failures and undetected errors.
Furthermore, these undesirable events may be correlated in time, thereby causing queue buildups at the source that induce unacceptable delays at the destination.
The latter issue is especially important for channels with memory, as correlation in service is known to exacerbate deviations in queueing systems.
This discussion points to the need to carefully explore the tradeoffs between queueing and coding for communication systems with tight delay requirements, giving due consideration to optimal block lengths and code rates.

Delay-sensitive systems have been studied in the past, leading to several landmark contributions~\cite{Ephremides-it98,Telatar-jsac95,Anantharam-it96}.
For example, the use of advanced power-control policies can be tailored to the needs of various applications~\cite{Berry-it02,Harsini-09,Bettesh-it06,Li-wcmc10}.
In many such articles, the emphasis is put on average delay and the optimization objective naturally leads to dynamic programming formulations~\cite{Fu-infocom03,Bettesh-it06,Cao-isit08,Steiner-it10}.
We stress that the wide applicability of Little's law can be leveraged to simplify the analysis of many systems where average delay is a prime consideration~\cite{Swamy-Asilomar08}.
On a different note, the recent advent of network coding and the complementary approach of channel coding over networks have been applied to short-block, delay-sensitive communications~\cite{Eryilmaz-it08,Cogill-it11}.
Such schemes seem especially well-suited for packet-loss networks, and the ensuing framework represents a potential alternative to automatic repeat requests when feedback is slow or error-prone.
Although closely related, these contributions differ from our formulation in that the main focus is on the operation of the system at the packet level, whereas we seek to characterize the impact of channel behavior at the symbol level.

Optimum code-rate selection has previously been studied for Gilbert-Elliott erasure channels with Bernoulli arrivals and maximum-likelihood decoding~\cite{ieee-tit-2013-pcpn}.
This prior line of work centers around random codes of fixed lengths, and it offers a distinct approach to assess the performance of communication systems operating over erasure channels.
The present article offers a significant extension to these existing contributions in that we examine finite-state error channels, we leverage pragmatic coding schemes and we adopt a scalable arrival profile.
The first important distinction between our findings and previously published results is the rigorous characterization of queueing behavior for communications over finite-state error channels, as opposed to erasure channels.
This is an important and nontrivial extension, which arises through the fact that erasure channels intrinsically pinpoint the location of channel distortion events at the receiver whereas error channels do not.
This lack of location information renders the decoding process much more challenging in the latter case.
Although technically more demanding, error channels permit the more realistic modeling of practical communication links.
For example, in our analysis, detected and undetected errors both demand appropriate considerations.

In addition to this channel enhancement, we leverage pragmatic coding schemes such as BCH encoding with bounded distance decoding to bring a pragmatic flavor to the analysis.
Furthermore, under random coding schemes, we present a novel framework to analyze overall system performance (e.g., probability of buffer overflow) using both optimal decoding and minimum distance decoding.
This perspective is very beneficial because the probability of decoding failure plays a crucial role in characterizing packet departures, queue transitions and the stationary behavior of the transmit buffer.

Another prime distinction between our current contribution and previous work is the adoption of a scalable arrival profile which is formed based on the Poisson process.
Among other advantages, the proposed framework allows for the rigorous comparison of coding schemes with different block lengths, something that could not be done before.
Indeed, this appears to be the first time one can perform the rigorous evaluation of queueing performance over block lengths.
By adopting a Poisson (or Markov modulated Poisson) model, we are able to overcome these limitations.
We emphasize that the scaling property of the Poisson process is crucial in enabling the fair comparison of systems with different block lengths.
The price to pay for this additional flexibility is a slightly more complicated analysis.
In particular, we need to employ an advanced version of the matrix-geometric method.
Our analysis leads to an enhanced framework for code design and resource allocation in the context of delay-sensitive wireless communications.

One of the challenges in dealing with block codes over finite-state channels with memory is the time dependencies among proximate decoding events.
For instance, if the underlying channel forms a Markov chain, then the decoding process becomes a hidden Markov process as block codes operate over series of channel states.
This often entails a difficult analysis of the queue behavior at the source.
To make this problem tractable, we use the idea of state augmentation which was also used in~\cite{ieee-tit-2013-pcpn}, where the value of the channel at the onset of a codeword is appended to the queue length.
Under this state augmentation, the coded system retains the Markov property, which facilitates the precise characterization of the queueing behavior at the transmitter.
This approach is paralleled in the present article, albeit in the general context of error channels.

We review and extend the necessary mathematical machinery to handle error events, as opposed to erasures, starting with the binary memoryless channel.
This step is pivotal in better understanding the encoding/decoding analysis of communication links with errors.
We then turn to finite-state channels with memory, as originally introduced by Gilbert~\cite{Gilbert-bell60} and Elliott~\cite{Elliott-bell63}.
We leverage the latter abstractions to assess how channel dependencies over time can affect overall performance.
It is well-known that correlation in service can significantly alter the behavior of a queueing system or network;
such changes should be expected in the present scenario as well.
Still, a novel facet of the problem we are considering is the study of how such dependencies affect the selection of optimal design parameters in terms of code rate and block length.
Furthermore, the framework presented in this paper can be used to derive novel and fundamental bounds on the maximum arrival rate that a wireless system can support when subject to certain quality of service requirements.

\section{Gilbert-Elliott Channel Model}
\label{section:ChannelModel}

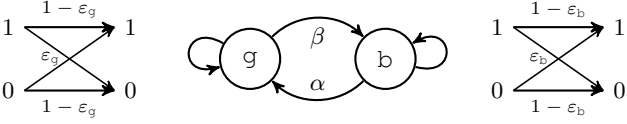
\begin{figure}[tb]
\begin{center}
\begin{tikzpicture}
[node distance = 12mm, draw=black, thick, >=stealth',
state/.style={circle, inner sep = 0pt, minimum size = 8mm, draw=black}]

\node[state] (l0) at (0,0) {\good};
\draw [->] (l0) to [out=150, in=195, looseness=5] (l0);

\node[state] (l1) at (1.8,0) {\bad}
  edge[<-, bend right=45] node[auto,swap,below] {\small{$\beta$}} (l0)
  edge[->, bend left=45] node[auto,above] {\small{$\alpha$}} (l0);
\draw [->] (l1) to [in=30,out=-15,looseness=5] (l1);

\node[coordinate] (l0In0) at (-3,0) [yshift=-12, label=left:{\small{$0$}}]{};
\node[coordinate] (l0Out0) [right= of l0In0, label=right:{\small{$0$}}]{}
  edge[<-] node[below] {\scriptsize{$1 - \varepsilon_\good$}} (l0In0);
\node[coordinate] (l0In1) at (-3,0) [yshift=12, label=left:{\small{$1$}}]{}
  edge[semithick, ->] node[left] {\scriptsize{$\varepsilon_\good$}} (l0Out0);
\node[coordinate] (l0Out1) [right= of l0In1, label=right:{\small{$1$}}]{}
  edge[<-] node[above] {\scriptsize{$1 - \varepsilon_\good$}} (l0In1)
  edge[semithick, <-] (l0In0);

\node[coordinate] (l1In0) at (3.5,0) [yshift=-12, label=left:{\small{$0$}}]{};
\node[coordinate] (l1Out0) [right= of l1In0, label=right:{\small{$0$}}]{}
  edge[<-] node[below] {\scriptsize{$1 - \varepsilon_\bad$}} (l1In0);
\node[coordinate] (l1In1) at (3.5,0) [yshift=12, label=left:{\small{$1$}}]{}
  edge[semithick, ->] node[left] {\scriptsize{$\varepsilon_\bad$}} (l1Out0);
\node[coordinate] (l1Out1) [right= of l1In1, label=right:{\small{$1$}}]{}
  edge[<-] node[above] {\scriptsize{$1 - \varepsilon_\bad$}} (l1In1)
  edge[semithick, <-] (l1In0);
\end{tikzpicture}
\end{center}
\caption{The Gilbert-Elliott model is one of the simplest non-trivial instantiations of a finite-state channel with memory.
State evolution over time forms a Markov chain, and the input-output relationship of this binary channel is governed by a state-dependent crossover probability, as illustrated above.}
\label{figure:A-Gilbert-Elliott-bit}
\end{figure}

At present, the term Gilbert-Elliott channel often refers to a wide class of finite-state fading channels that model communication links with memory.
In our article, however, we allude to its original definition and we use the denomination Gilbert-Elliott channel to designate a binary symmetric channel that features two possible states: a \emph{good} state $\good$ with crossover probability $\varepsilon_{\good}$, and a \emph{bad} state $\bad$ with crossover probability $\varepsilon_{\bad}$.
While simple, this model can account for uncertainties associated with transmitting symbols over a noisy channel and correlation over time.
The evolution of the channel is governed by a finite-state Markov chain.
We denote the transition probability from $\bad$ to $\good$ by $\alpha$, and we label the transition probability in the reverse direction by $\beta$.
Thus, the channel evolution forms a Markov chain with transition probability matrix
\begin{equation*}
\renewcommand{\arraystretch}{0.5}
\mathbf{P}= \begin{bmatrix} 1-\alpha & \alpha \\
\beta & 1-\beta \end{bmatrix} .
\end{equation*}
A graphical representation of this channel appears in Fig.~\ref{figure:A-Gilbert-Elliott-bit}.
It is worth mentioning that the steady-state probabilities of the good and bad states are $\frac{\alpha}{\alpha+\beta}$ and $\frac{\beta}{\alpha+\beta}$, respectively.

We note that, in defining the matrix $\mathbf{P}$, we have implicitly ordered the states from bad to good.
With a slight abuse of notation, we use this bijection between channel states and their numerical indices to refer to specific entries in the matrix.
We employ random variable {$C_{n}$} to denote the state of the channel at time $n$.
Then, entry $[\mathbf{P}]_{c,d}$ represents the probability of a channel transition to state $d$, given that the current state is $c$.
For groups of random variables, we use the common expression $P_{\cdot|\cdot}(\cdot|\cdot)$ to denote conditional joint probability mass functions.
Accordingly, we can write $P_{C_{n+1}|C_{n}} (d | c) = \Pr (C_{n+1} = d | C_n = c)$, where $c,d \in \{ \bad, \good \}$.
In a similar fashion, $P_{C_{n+N}|C_{n}}(d | c)$ can be obtained by looking at the proper entry of matrix $\mathbf{P}^N$.

To proceed, we need a way to compute the conditional distribution of the number of errors that occur during $N$ consecutive uses of the channel.
Let $E$ denote the number of errors occurring in a data block.
The distribution of $E$ can be obtained using the matrix of polynomials
\begin{equation*}
\renewcommand{\arraystretch}{1.0}
\mathbf{P}_{x}
= \begin{bmatrix}
(1 - \alpha)(1 - \varepsilon_{\bad} + \varepsilon_{\bad} x)
& \alpha(1 - \varepsilon_{\bad} + \varepsilon_{\bad} x) \\
\beta (1 - \varepsilon_{\good} + \varepsilon_{\good} x)
& (1 - \beta)(1 - \varepsilon_{\good} + \varepsilon_{\good} x)
\end{bmatrix}.
\end{equation*}
Throughout, we employ {$\llbracket x^{j}\rrbracket$} to represent the linear functional that maps a polynomial in {$x$} to the coefficient of {$x^j$}.
Using this notation, we get $P_{E,C_{N+1}|C_1}(e,d|c) = \Pr(E=e,C_{N+1}=d | C_1=c) = \llbracket x^{e} \rrbracket \left[\mathbf{P}_{x}^{N}\right]_{c,d}$.
Eventually, we will use this distribution to compute the conditional probabilities of decoding failure and undetected error.
We note that closed-form recursions for these values have been derived a number of times in the past \cite{Elliott-bell63,Wilhelmsson-com99}.

\section{Arrivals, Departures, and Feedback}
\label{section:ArrivalsDepartures}

\begin{figure}
\centering
\ifoneCol
\scalebox{.9}{\input{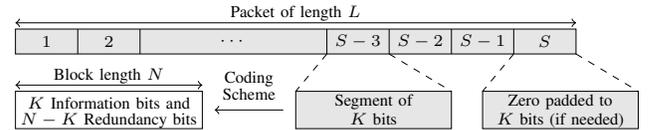}}
\else
\scalebox{.9}{\begin{tikzpicture}[scale=.92]

\draw [<->] (0,0.5)--(9,0.5) ;
\node [above] at (4.5,0.5-0.1) {\scriptsize Packet of length $L$} ;
\draw [fill=gray!20] (0,0) rectangle (9,0.4);
\draw (1,0)--(1,0.4) ; \draw (2,0)--(2,0.4) ;
\draw (5,0)--(5,0.4) ; \draw (6,0)--(6,0.4) ; \draw (7,0)--(7,0.4) ; \draw (8,0)--(8,0.4) ;
\node at (0.5,0.2) {\scriptsize $1$} ;
\node at (1.5,0.2) {\scriptsize $2$} ;
\node at (3.5,0.2) {\scriptsize $\cdots$} ;
\node at (5.5,0.2) {\scriptsize $S-3$} ;
\node at (6.5,0.2) {\scriptsize $S-2$} ;
\node at (7.5,0.2) {\scriptsize $S-1$} ;
\node at (8.5,0.2) {\scriptsize $S$} ;
\newcommand{\myy}{-1.2}
\draw [fill=gray!20] (7.5,\myy) rectangle (10,\myy+0.6) ;
\node [above] at (8.75,\myy+0.3-0.12) {\scriptsize Zero padded to } ;
\node [below] at (8.75,\myy+0.3+0.08) {\scriptsize $K$ bits (if needed)} ;
\draw [fill=gray!20] (4.5,\myy) rectangle (7.0,\myy+0.6) ;
\node [above] at (5.75,\myy+0.3-0.12) {\scriptsize Segment of} ;
\node [below] at (5.75,\myy+0.3+0.08) {\scriptsize $K$ bits} ;
\draw [<->] (0,\myy+0.6+0.1)--(3.0,\myy+0.6+0.1) ;
\node [above] at (1.5,\myy+0.6+0.1-0.1) {\scriptsize Block length $N$} ;
\draw [fill=white] (0,\myy) rectangle (3.0,\myy+0.6) ;
\node [above] at (1.5,\myy+0.3-0.12) {\scriptsize $K$ Information bits and} ;
\node [below] at (1.5,\myy+0.3+0.08) {\scriptsize $N-K$ Redundancy bits} ;
\draw [<-] (3.0+0.2,\myy+0.3)--(4.5-0.2,\myy+0.3) ;
\node [above] at (3.75,\myy+0.55) {\scriptsize Coding} ;
\node         at (3.75,\myy+0.55) {\scriptsize Scheme} ;
\draw [dashed] (9,0)--(10,\myy+0.6) ;
\draw [dashed] (8,0)--(7.5,\myy+0.6) ;
\draw [dashed] (6,0)--(7.0,\myy+0.6) ;
\draw [dashed] (5,0)--(4.5,\myy+0.6) ;
\end{tikzpicture} }
\fi
\caption{Each packet is divided into $S$ segments, and a channel encoding scheme is used to encode each segment}
\label{fig:Segmentation-of-a}
\end{figure}

\begin{figure}
\centering
\ifoneCol
\scalebox{.9}{\input{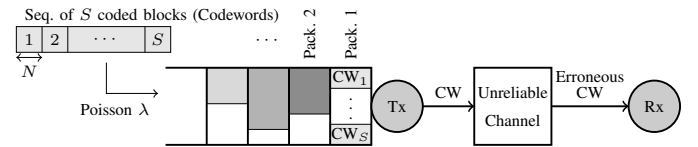}}
\else
\scalebox{.9}{\begin{tikzpicture}[scale=0.76]

\newcommand{\myy}{1.8} ;
\node [right] at (0,\myy+0.5+0.2) {\scriptsize Seq. of $S$ coded blocks (Codewords)} ;
\draw [fill=gray!20](0,\myy+0) rectangle (3.0,\myy+0.5) ;
\draw (0.5,\myy+0)--(0.5,\myy+0.5) ; \draw (1.0,\myy+0)--(1.0,\myy+0.5) ; \draw (2.5,\myy+0)--(2.5,\myy+0.5) ;
\node at (0.25,\myy+0.25) {\scriptsize $1$} ;
\node at (0.75,\myy+0.25) {\scriptsize $2$} ;
\node at (1.75,\myy+0.25) {\scriptsize $\cdots$} ;
\node at (2.75,\myy+0.25) {\scriptsize $S$} ;
\draw [<->] (0,\myy-0.1)--(0.5,\myy-0.1) ;
\node [below] at (0.25,\myy-0.1) {\scriptsize $N$} ;
\newcommand{\myx}{2.9}
\renewcommand{\myy}{0}
\draw [fill=gray!30] (\myx+0.8,\myy+0.8) rectangle (\myx+1.6,\myy+1.5) ;
\draw [fill=gray!60] (\myx+1.6,\myy+0.3) rectangle (\myx+2.4,\myy+1.5) ;
\draw [fill=gray!90] (\myx+2.4,\myy+0.6) rectangle (\myx+3.2,\myy+1.5) ;
\draw [fill=gray!20] (\myx+3.2,\myy+1.1) rectangle (\myx+4.0,\myy+1.5) ;
\node at (\myx+3.6,\myy+1.3) {\scriptsize $\mbox{CW}_1$} ;
\node at (\myx+3.6,\myy+0.85) {\scriptsize $\vdots$} ;
\draw [fill=gray!20] (\myx+3.2,\myy+0.0) rectangle (\myx+4.0,\myy+0.4) ;
\node at (\myx+3.6,\myy+0.2) {\scriptsize $\mbox{CW}_S$} ;
\node [rotate=90,right] at (\myx+3.6,\myy+1.5) {\scriptsize Pack. 1} ;
\node [rotate=90,right] at (\myx+2.8,\myy+1.5) {\scriptsize Pack. 2} ;
\node [above]           at (\myx+2.0,\myy+1.8) {\scriptsize $\cdots$} ;
\draw [thick] (\myx+0.0,\myy+0)--(\myx+4.0,\myy+0)--(\myx+4.0,\myy+1.5)--(\myx+0,\myy+1.5) ;
\draw [thick] (\myx+0.8,\myy+0)--(\myx+0.8,\myy+1.5) ;
\draw [thick] (\myx+1.6,\myy+0)--(\myx+1.6,\myy+1.5) ;
\draw [thick] (\myx+2.4,\myy+0)--(\myx+2.4,\myy+1.5) ;
\draw [thick] (\myx+3.2,\myy+0)--(\myx+3.2,\myy+1.5) ;
\draw [thick,fill=gray!40] (\myx+4.5,\myy+0.75) circle (0.5) ;
\draw [thick,->] (\myx+5.0,\myy+0.75)--(\myx+6.0,\myy+0.75) ;
\node [above] at (\myx+5.5,\myy+0.75) {\scriptsize CW} ;
\node at (\myx+4.5,\myy+0.75) {\scriptsize Tx} ;
\draw [thick,fill=white](\myx+6,\myy+0) rectangle (\myx+7.5,\myy+1.5) ;
\draw [thick,->] (\myx+7.5,\myy+0.75)--(\myx+9.0,\myy+0.75) ;
\node [above] at (\myx+8.25,\myy+0.75+0.3) {\scriptsize Erroneous} ;
\node [above] at (\myx+8.25,\myy+0.75) {\scriptsize CW} ;
\node [above] at (\myx+6.75,\myy+0.75) {\scriptsize Unreliable} ;
\node [below] at (\myx+6.75,\myy+0.75) {\scriptsize Channel} ;
\draw [thick,fill=gray!40] (\myx+9.5,\myy+0.75) circle (0.5) ;
\node at (\myx+9.5,\myy+0.75) {\scriptsize Rx} ;
\draw [->] (1.75,1.8-0.2)--(1.75,1.0)--(2.75,1.0) ;
\node [left] at (2.75,1.0-0.3) {\scriptsize Poisson $\lambda$} ;
\end{tikzpicture}}
\fi
\caption{Coded segments are transmitted over the unreliable communication link. A data packet is discarded from the transmit buffer only when all its codewords are successfully transmitted}
\label{fig:Transmission-of-the}
\end{figure}

In this section, we describe the elements that compose our queueing system.
Suppose that a packet of length $L$ needs to be sent over the Gilbert-Elliott channel to a destination.
In the proposed framework, this packet is divided into $S$ segments, each containing $K$ information bits.
The last segment is zero padded, if needed, to conform to the prescribed length.
A BCH code or a random code is used to encode each data segment into a codeword of length $N$ (see Fig.~\ref{fig:Segmentation-of-a}).
These codewords are then transmitted over the communication link.
Packet arrivals at the source are initially assumed to form an instance of a Poisson process with rate $\lambda$ packets per channel use (see Fig.~\ref{fig:Transmission-of-the}).
Therefore, the number of packets expected to arrive during an interval of length $N$ is equal to $\lambda_N = \lambda N$.
As we will see, our framework can accommodate more general packet arrivals, such as Markov processes with discrete state spaces \cite{Das-icupc97,Kang-globecom95}.
This comes at the expense of additional bookkeeping.
For instance, a Markov modulated Poisson process (MMPP) with distinct arrival rates, can be employed to better capture bursty traffic~\cite{Lee-vt10} and fluctuations in workload.

Packet sizes are assumed to form a sequence of independent and identically distributed random variables, where each element has a geometric distribution with parameter $\rho \in (0,1)$.
Mathematically, we write $\Pr(L = \ell) = (1 - \rho)^{\ell-1} \rho$ where $\ell \geq 1$.
This assumption plays a key role in our article and it has been selected, partly, to facilitate the analysis we wish to carry.
In particular, the memoryless property of the geometric distribution makes for a tractable queueing model.
Not too surprisingly, having a geometric distribution for the size of packets is commonplace in the literature \cite{Sriram-jsac86,Heffes-jsac86}.

We can further relate the packet-length distribution to the progression of coded transmissions.
For fixed block length $N$ and code rate $R$, every successful decoding event reveals exactly $RN$ information bits to the destination.
As such, when a data packet contains $L$ bits, one needs to successfully decode $S=\left\lceil \frac{L}{RN}\right\rceil$ codewords to complete the transmission of the entire packet.
We note that random variable $S$ possesses a geometric distribution with
$\Pr(S=s) = (1 - \rho_r)^{s-1} \rho_r$, where $s \geq 1$ and $\rho_r = 1 - (1 - \rho)^{RN}$.
Thus, in the current setting, the number of coded blocks per data packet $S$ retains the memoryless property.
We emphasize that, in our framework, a data packet is discarded from the transmit buffer if and only if the destination acknowledges reception of the latest codeword and this codeword contains the last parcel of information corresponding to the head packet.
The departure process is governed by the parameters of the channel and the coding scheme adopted.
Generally, a lower code rate yields smaller probabilities of decoding failure, but it also entails having more data segments to send.
Thus, for a given channel and load, it is important to choose the block length and the code rate which give the best overall performance.
In Section~\ref{sec:Numerical-Results}, we present simulation results for a system with packets of a constant size and we compare its performance to the corresponding system with geometrically distributed packet lengths.

A subtle, yet important aspect associated with automatic repeat request over unreliable connections is the amount of feedback needed by a particular scheme.
Using shorter block lengths necessarily entails more frequent feedback messages from the destination.
In general, adequately evaluating the costs and benefits of various feedback strategies is a complicated task.
Since this is not a prime objective of this article, we circumvent this issue by making simple assumptions.
We assume that feedback is instantaneously and faithfully received at the source; this idealized view is frequently found in the literature \cite{Fantacci-vt96,lugand-com89}.
In contrast, any detailed analysis of feedback requires making strong assumptions about correlation between the forward and reverse links, the delay associated with receiving feedback, and mechanisms to cope with corrupted messages.
Although these issues warrant attention, they are outside the scope of this article.
Beyond that, we hypothesize that the price of feedback is captured by having a portion of every data segment dedicated to a header of length~$h$.
Of course, this reduces the size of the packet payload to $RN - h$.
This crude approximation treats feedback bits as constant overhead, and it is a modest step in better accounting for control messages.
Feedback overhead will affect the number of segments contained in a data packet.
If $h$ information bits in every segment pertain to the header, then the number of successful codeword transmissions necessary to transfer a packet becomes $S = \left\lceil \frac{L}{K-h} \right\rceil$, a slight variation compared to the original value.
Nevertheless, $S$ retains a geometric distribution, albeit with parameter $\rho_r = 1 - (1-\rho)^{K-h}$.

A very important aspect of queueing systems is stability.
The Foster-Lyapunov criterion ensures that our simple system remains stable so long as the packet service rate exceeds the arrival rate.
To calculate the mean service rate, we recall that a packet leaves the queue whenever a codeword is decoded successfully and this codeword carries the last data segment of the head packet.
Let $P_{\mathrm{s}|E}(e)$ and $P_{\mathrm{f}|E}(e)$ denote the conditional probabilities of decoding success and failure, respectively, given the number of errors within a block, $E = e$.
By reciprocity, the conditional success probability is equal to $P_{\mathrm{s}|E}(e) = 1 - P_{\mathrm{f}|E}(e)$.
Then, the average service rate can be computed as $\mu_N = \rho_r \mathrm{E} \left[ P_{\mathrm{s}|E}(e) \right]$ packets per codeword transmission.
The stability factor for this system is $\frac{\lambda_N}{\mu_N}$, and the process is stable provided that this ratio is less than unity.
Conditional failure probabilities will be computed explicitly in Section~\ref{section:fail} for different channels and coding schemes.

\section{Queueing Model}
\label{section:QueueingModel}

We are ready to examine more closely the queueing behavior of our communication link.
Throughout, we use $Q_s$ to denote the number of packets waiting in the transmit buffer.
The channel state at the same instant is $C_{sN+1}$.
By grouping these two random variables together, we can construct a discrete-time Markov chain (DTMC), which we write $U_s = (C_{sN+1}, Q_s)$.
The resulting DTMC is of the M/G/1 type, and there are many established techniques that apply to such systems \cite{Riska-sigmet02,Karlin-75,ieee-tit-2013-pcpn}.
We note that for the binary symmetric channel, input-output properties are unchanged over time.
In this degenerate case, the queue length $Q_s$ contains all the information relevant to the DTMC, and the random variable $U_s$ is mathematically equivalent to the state of the transmit buffer.

Using the total probability theorem, the transition probabilities for the DTMC $\{ U_s \}$ can be decomposed as
\begin{align*}
&\Pr(U_{s+1} = (d, q_{s+1}) | U_s = (c, q_s)) \\
&= \textstyle\sum_{e \in \mathbb{N}_0} P_{Q_{s+1}|E, Q_{s}} (q_{s+1} | e, q_s)
P_{E,C_{(s+1)N+1} | C_{sN+1}} (e, d | c) .
\end{align*}
Examining the summands, we need to derive expressions for $P_{Q_{s+1} | E, Q_s} (q_{s+1} | e, q_s)$.
Suppose that the current number of packets in the queue is $Q_s = q_s$.
Then, admissible values for $Q_{s+1}$ are restricted to the collection $\{ q_s-1, q_s, q_s+1, \ldots\}$.
The corresponding transition probabilities are given by
\begin{align}
&P_{Q_{s+1} | E, Q_s} (q_s - 1 | e, q_s)
= a_0 (1 - P_{\mathrm{f}|E}(e)) \rho_r, \nonumber\\
&P_{Q_{s+1} | E, Q_s} (q_s + i | e, q_s)
= a_{i+1} (1 - P_{\mathrm{f}|E}(e)) \rho_r \nonumber\\
&\qquad+ a_i \left( P_{\mathrm{f}|E}(e)
+ (1 - P_{\mathrm{f}|E}(e))(1 - \rho_r) \right) ,
\quad i \geq 0 \label{eq:transProb}
\end{align}
where $a_i = \frac{(\lambda N)^{i}}{i!} e^{-\lambda N}$ is the probability that $i$ packets arrive during the transmission of one codeword.
When the queue is empty, $\{Q_{s}=0\}$, the transition probabilities reduce to $P_{Q_{s+1} | E, Q_s} (q_s + i | e, 0) = a_i$ with $i \geq 0$.

Using these equations, we can get the probability transition matrix of the Markov process $\{U_{s}\}$.
First, we introduce the following convenient notation, where $q \in \mathbb{N}_0$ and $c,d \in \{\good,\bad\}$,
\begin{align*}
&\mu_{cd}^i = \Pr(U_{s+1} = (d, q+i) | U_s = (c, q)) \quad i \geq 1,\\
&\kappa_{cd} = \Pr(U_{s+1} = (d, q) | U_s = (c,q)) \\
&\xi_{cd} = \Pr(U_{s+1} = (d, q-1) | U_s = (c,q)) .
\end{align*}
Similarly, when the queue is empty, we write $\mu_{cd}^{i0} = \Pr(U_{s+1} = (d, i) | U_s = (c, 0))$ and $\kappa_{cd}^0 = \Pr(U_{s+1} = (d,0) | U_s = (c,0))$.
Possible state transitions are illustrated in Fig.~\ref{fig:State-space-and}.

\begin{figure}[!t]
\centering
\scalebox{0.48}{
%
%
%
%
%
%
\begin{tikzpicture}[->,>=stealth',shorten >=1pt,auto,node distance=2.8cm]
\tikzstyle{every state}=[fill=white,draw=black,thick,text=black,scale=1]
\node[state]         (A)              {$(\good,0)$};
\node[state]         (B) [right of=A] {$(\good,1)$};
\node[state]         (C) [right of=B] {$(\good,2)$};
\node[state]         (D) [right of=C] {$(\good,3)$};
\node (I) [right of=D] {$\cdots$};
\path (A) edge  [bend left=20] node[above] {} (I);
\path (B) edge  [bend left=20] node[above] {} (I);
\path (C) edge  [bend left=20] node[above] {} (I);
\path (D) edge  [bend left=10,dashed] node[above] {} (I);
\path (I) edge  [bend left=10,dashed] node[above] {} (D);
\path (A) edge  [bend left=10] node[above] {} (B);
\path (B) edge  [bend left=10] node[below] {} (A);
\path (B) edge  [bend left=10] node[above] {} (C);
\path (C) edge  [bend left=10] node[below] {} (B);
\path (C) edge  [bend left=10] node[above] {} (D);
\path (D) edge  [bend left=10] node[below] {} (C);
\path (A) edge  [bend left=20] node[above] {} (C);
\path (A) edge  [bend left=20] node[above] {} (D);
\path (B) edge  [bend left=20] node[above] {} (D);
\tikzstyle{every state}=[fill=white,draw=black,thick,text=black,scale=1]
\node[state]         (E) [below of=A] {$(\bad,0)$};
\path (A) edge  [bend right=10] node[left] {} (E);
\path (E) edge  [bend right=10] node[right] {} (A);
\tikzstyle{every state}=[fill=white,draw=black,thick,text=black,scale=1]
\node[state]         (F) [below of=B] {$(\bad,1)$};
\node[state]         (G) [below of=C] {$(\bad,2)$};
\node[state]         (H) [below of=D] {$(\bad,3)$};
\path (B) edge  [bend right=10] node[right] {} (E);
\path (E) edge  [bend right=10] node[left] {} (B);
\path (E) edge  [bend right=10] node[left] {} (C);
\path (E) edge  [bend right=5] node[left] {} (D);
\path (A) edge  [bend right=10] node[right] {} (F);
\path (A) edge  [bend right=-10] node[right] {} (G);
\path (A) edge  [bend right=-5] node[right] {} (H);
\path (F) edge  [bend right=10] node[left] {} (A);
\path (B) edge  [bend right=10] node[right] {} (G);
\path (B) edge  [bend right=-10] node[right] {} (H);
\path (G) edge  [bend right=10] node[left] {} (B);
\path (C) edge  [bend right=10] node[right] {} (F);
\path (F) edge  [bend right=10] node[left] {} (C);
\path (F) edge  [bend right=10] node[left] {} (D);
\path (C) edge  [bend right=10] node[right] {} (H);
\path (H) edge  [bend right=10] node[left] {} (C);
\path (D) edge  [bend right=10] node[right] {} (G);
\path (G) edge  [bend right=10] node[left] {} (D);
\path (I) edge  [bend right=10,dashed] node[right] {} (H);
\path (H) edge  [bend right=10,dashed] node[left] {} (I);
\path (B) edge  [bend right=10] node[left] {} (F);
\path (F) edge  [bend right=10] node[right] {} (B);
\path (C) edge  [bend right=10] node[left] {} (G);
\path (G) edge  [bend right=10] node[right] {} (C);
\path (D) edge  [bend right=10] node[left] {} (H);
\path (H) edge  [bend right=10] node[right] {} (D);

\path (E) edge  [bend left=10] node[above] {} (F);
\path (F) edge  [bend left=10] node[below] {} (E);
\path (F) edge  [bend left=10] node[above] {} (G);
\path (G) edge  [bend left=10] node[below] {} (F);
\path (G) edge  [bend left=10] node[above] {} (H);
\path (H) edge  [bend left=10] node[below] {} (G);

\path (E) edge  [bend right=20] node[above] {} (G);
\path (E) edge  [bend right=20] node[above] {} (H);
\path (F) edge  [bend right=20] node[above] {} (H);

\node (J) [right of=H] {$\cdots$};
\path (E) edge  [bend right=20] node[above] {} (J);
\path (F) edge  [bend right=20] node[above] {} (J);
\path (G) edge  [bend right=20] node[above] {} (J);
\path (H) edge  [bend right=10,dashed] node[above] {} (J);
\path (J) edge  [bend right=10,dashed] node[above] {} (H);
\path (D) edge  [bend right=10,dashed] node[right] {} (J);
\path (J) edge  [bend right=10,dashed] node[left] {} (D);
\end{tikzpicture}
}
\caption{\label{fig:State-space-and}State space and transition diagram for
the aggregate queued process {$\{U_{s}\}$}; self-transitions are
intentionally omitted.}
\ifoneCol
\else
\vspace{\floatsep}
\scalebox{0.61}{\begin{tikzpicture}[->,>=stealth',shorten >=1pt,auto,node distance=0.6cm]
    \node (State00) [state2] {$\phantom{\!+\!}(\good,0)\phantom{1}$};
    \node (State01) [state2,below=of State00] {$\phantom{\!+\!}(\bad,0)\phantom{1}$} ;

    \begin{pgfonlayer}{background}

        \path (State00.west |- State00.north)+(-0.2,0.1) node (a) {};
        \path (State01.east |- State01.south)+(0.2,-0.1) node (b) {};
        \path (State00.east |- State00.north)+(-0.3,0.1) node (c) {};
        \path (State00.east |- State00.north)+(-0.7,0.7) node (po1) {\LARGE{$\color{gray}\mathbf{\pi}_0$}};\normalsize
        \path (State01.east |- State01.south)+(-0.7,-0.1) node (cc) {};
        \path [fill=gray!20,rounded corners, draw=black] (a) rectangle (b);

    \end{pgfonlayer}

      \node (State02) [state2,right=of State00] {$\phantom{\!+\!}(\good,1)\phantom{1}$};
      \node (State03) [state2,below=of State02] {$\phantom{\!+\!}(\bad,1)\phantom{1}$} ;

    \begin{pgfonlayer}{background}

        \path (State02.west |- State02.north)+(-0.2,0.1) node (d) {};
        \path (State03.east |- State03.south)+(0.2,-0.1) node (e) {};
        \path (State02.east |- State02.north)+(-0.3,0.1) node (f) {};
        \path (State02.east |- State02.north)+(-0.7,0.7) node (po2) {}; 
        \path (State03.east |- State03.south)+(-0.7,-0.1) node (ff) {};
        \path [fill=gray!20,rounded corners, draw=black] (d) rectangle (e);

    \end{pgfonlayer}

        \node (S2) [right=of State02] {};
        \node (I2)  [below=of S2]{\LARGE$\cdots$};\normalsize
        \node (u2) [below=of I2] {};
    \path (c) edge  [bend left=20] node[above=-3pt] {$\hat{\mathbf{F}}^{(1)}$} (f);
    \begin{pgfonlayer}{background}
        \path [fill=white,rounded corners, draw=white] (S2) rectangle (u2);
    \end{pgfonlayer}

        \node (State04) [state2,right=of S2] {$\phantom{\!+\!}(\good,i)\phantom{1}$};
        \node (State05) [state2,below=of State04] {$\phantom{\!+\!}(\bad,i)\phantom{1}$} ;

    \begin{pgfonlayer}{background}

        \path (State04.west |- State04.north)+(-0.2,0.1) node (g) {};
        \path (State05.east |- State05.south)+(0.2,-0.1) node (h) {};
        \path (State04.east |- State04.north)+(-0.3,0.1) node (i) {};
        \path (State04.east |- State04.north)+(-0.7,0.7) node (po3) {\LARGE{$\color{gray}\mathbf{\pi}_i$}};\normalsize
        \path (State05.east |- State05.south)+(-0.7,-0.1) node (ii) {};
        \path [fill=gray!20,rounded corners, draw=black] (g) rectangle (h);

    \end{pgfonlayer}
    \path (c) edge  [bend left=40] node[above=-3pt] {$\hat{\mathbf{F}}^{(i)}$} (i);
    \path (f) edge  [bend left=20] node[above=-3pt] {$\mathbf{F}^{(i-1)}$} (i);
    \path (ff) edge  [bend left=20] node[below] {$\mathbf{B}$} (cc);

        \node (State06) [state2,right=of State04] {$(\!\good,\!i\!+\!1)$};
        \node (State07) [state2,below=of State06] {$(\!\bad,\!i\!+\!1)$} ;

    \begin{pgfonlayer}{background}

        \path (State06.west |- State06.north)+(-0.2,0.1) node (j) {};
        \path (State07.east |- State07.south)+(0.2,-0.1) node (k) {};
        \path (State06.east |- State06.north)+(-0.3,0.1) node (l) {};
        \path (State07.east |- State07.south)+(-0.7,-0.1) node (ll) {};
        \path [fill=gray!20,rounded corners, draw=black] (j) rectangle (k);

    \end{pgfonlayer}
    \path (c) edge  [bend left=40] node[above=-3pt] {$\hat{\mathbf{F}}^{(i+1)}$} (l);
    \path (f) edge  [bend left=40] node[above=-3pt] {$\mathbf{F}^{(i)}$} (l);
    \path (i) edge  [bend left=20] node[above=-3pt] {$\mathbf{F}^{(1)}$} (l);
    \path (ll) edge  [bend left=20] node[below] {$\mathbf{B}$} (ii);

        \node (State08) [state2,right=of State06] {$(\!\good,\!i\!+\!2)$};
        \node (State09) [state2,below=of State08] {$(\!\bad,\!i\!+\!2)$} ;

    \begin{pgfonlayer}{background}

        \path (State08.west |- State08.north)+(-0.2,0.1) node (m) {};
        \path (State09.east |- State09.south)+(0.2,-0.1) node (n) {};
        \path (State08.east |- State08.north)+(-0.3,0.1) node (o) {};
        \path (State08.east |- State08.north)+(-0.5,0.7) node (po5) {\LARGE{$\color{gray}\mathbf{\pi}_{i+2}$}};\normalsize
        \path (State09.east |- State09.south)+(-0.7,-0.1) node (O) {};
        \path [fill=gray!20,rounded corners, draw=black] (m) rectangle (n);

    \end{pgfonlayer}
        \path (State09) edge  [bend left=10, very thick,draw=gray!100] node[near start=1pt,above] {\large$\color{gray!100}\mathbf{G}^2[2,1]$} (State04);\normalsize
        \path (State06) edge  [bend left, very thick,draw=gray!100] node[left] {\large$\color{gray!100}\mathbf{G}[1,2]$} (State05);\normalsize
        \path (State05) edge  [bend left, very thick,draw=gray!100] node[near start=2pt,below=-1pt] {\large$\color{gray!100}\mathbf{G}^{i-1}[2,1]$} (State02);\normalsize
    \path (c) edge  [bend left=40] node[above=-3pt] {$\hat{\mathbf{F}}^{(i+2)}$} (o);
    \path (f) edge  [bend left=40] node[above=-3pt] {$\mathbf{F}^{(i+1)}$} (o);
    \path (i) edge  [bend left=40] node[above=-3pt] {$\mathbf{F}^{(2)}$} (o);
    \path (l) edge  [bend left=20] node[above=-3pt] {$\mathbf{F}^{(1)}$} (o);
    \path (O) edge  [bend left=20] node[below] {$\mathbf{B}$} (ll);

    \path (ll) edge  [loop below,looseness=8,out=-60,in=-120] node[left] {$\mathbf{A}$} (ll);
    \path (O) edge   [loop below,looseness=8,out=-60,in=-120] node[left] {$\mathbf{A}$} (O);
    \path (ii) edge  [loop below,looseness=8,out=-60,in=-120] node[left] {$\mathbf{A}$} (ii);
    \path (ff) edge  [loop below,looseness=8,out=-60,in=-120] node[left] {$\mathbf{A}$} (ff);
    \path (cc) edge  [loop below,looseness=8,out=-60,in=-120] node[left] {$\hat{\mathbf{A}}$} (cc);
\node (S3) [right=of State08] {};
        \node (I3)  [below=of S3]{\LARGE$\cdots$};\normalsize
        \node (u3) [below=of I3] {};

    \begin{pgfonlayer}{background}
        \path [fill=white,rounded corners, draw=white, dashed] (S3) rectangle (u3);
    \end{pgfonlayer}
    \path (State08.east |- State08.north)+(0.1,1.1) node (o) {\LARGE$\cdots$};\normalsize
\end{tikzpicture}
}
\caption{\label{fig:G}Level transition diagram and probabilistic interpretation of $\mathbf{G}$}
\fi
\end{figure}

Next, we review briefly the matrix-geometric method, an efficient way to compute the stationary distributions of chains with repetitive structures.
We can represent the equilibrium distribution of our system as a semi-infinite vector $\boldsymbol{\pi} = (\pi(1), \pi(2), \ldots)$, where $\pi(2q+1) = \Pr(C=\bad,Q=q)$ and $\pi(2q+2) = \Pr(C=\good,Q=q)$.
Alternatively, we can group pairs of states together and write $\boldsymbol{\pi} = [ \boldsymbol{\pi}_0 \; \boldsymbol{\pi}_1 \; \boldsymbol{\pi}_2 \; \cdots ]$ where $\boldsymbol{\pi}_q$ comprises the stationary probabilities of the $q$th level of the chain with $\boldsymbol{\pi}_q = [\pi(2q+1) \; \pi(2q+2)]$.
Using this notation, one can express the detailed balance equation $\boldsymbol{\pi} \mathbf{T} = \boldsymbol{\pi}$ in terms of the transition probability matrix $\mathbf{T}$, which appears in block-partitioned form below
\begin{equation*}
\renewcommand{\arraystretch}{0.4}
\mathbf{T} = \left[ \begin{array}{ccccc}
\hat{\mathbf{A}} & \hat{\mathbf{F}}^{(1)}
& \hat{\mathbf{F}}^{(2)} & \hat{\mathbf{F}}^{(3)} & \cdots \\
\mathbf{B} & \mathbf{A} & \mathbf{F}^{(1)} & \mathbf{F}^{(2)} & \cdots \\
\mathbf{0} & \mathbf{B} & \mathbf{A} & \mathbf{F}^{(1)} & \cdots \\
\vdots & \vdots & \vdots & \vdots & \ddots
\end{array}\right] .
\end{equation*}
The labels $\mathbf{A}$, $\mathbf{F}$, and $\mathbf{B}$ symbolize local, forward, and backward transition-probability blocks, respectively;
the superscript $(i)$ indicates that $i$ additional data packets are present in the buffer at the next time instant;
and the hat designates instances where the queue is initially empty.
More specifically, we have
\begin{equation*}
\mathbf{F}^{(i)} =
\begin{bmatrix} \mu_{\mathrm{bb}}^{i} & \mu_{\mathrm{bg}}^{i} \\
\mu_{\mathrm{gb}}^{i} & \mu_{\mathrm{gg}}^{i} \end{bmatrix}, \;
\mathbf{A} =
\begin{bmatrix} \kappa_{\mathrm{bb}} & \kappa_{\mathrm{bg}} \\
\kappa_{\mathrm{gb}} & \kappa_{\mathrm{gg}} \end{bmatrix}, \;
\mathbf{B} =
\begin{bmatrix} \xi_{\mathrm{bb}} & \xi_{\mathrm{bg}} \\
\xi_{\mathrm{gb}} & \xi_{\mathrm{gg}} \end{bmatrix} .
\end{equation*}
For an empty queue, the blocks are
\begin{xalignat*}{2}
\hat{\mathbf{F}}^{(i)} &=
\begin{bmatrix} \mu_{\mathrm{bb}}^{i0} & \mu_{\mathrm{bg}}^{i0} \\
\mu_{\mathrm{gb}}^{i0} & \mu_{\mathrm{gg}}^{i0} \end{bmatrix} &
\hat{\mathbf{A}} &=
\begin{bmatrix} \kappa_{\mathrm{bb}}^{0} & \kappa_{\mathrm{bg}}^{0} \\
\kappa_{\mathrm{gb}}^{0} & \kappa_{\mathrm{gg}}^{0} \end{bmatrix} .
\end{xalignat*}
Figure~\ref{fig:G} shows the possible transitions among the different levels of the system.

\begin{proposition}
Let $\mathbf{G}$ be the limiting matrix of the recursion
\begin{equation} \label{eq:G}
\mathbf{G}_{i+1} = -\mathbf{L}^{-1} ( \mathbf{B}
+ \textstyle\sum_{j=1}^{\infty} \mathbf{F}^{(j)} \mathbf{G}_i^{j+1} )
\end{equation}
starting from $\mathbf{G}_0 = \mathbf{0}$ and where $\mathbf{L} = \mathbf{A} - \mathbf{I}$.
For $j \geq 1$, the stationary probability vectors $\boldsymbol{\pi}_j$ associated with $\mathbf{T}$ are given by
\begin{equation*}
\boldsymbol{\pi}_j = - ( \boldsymbol{\pi}_0 \hat{\mathbf{S}}^{(j)}
+ \textstyle\sum_{k=1}^{j-1} \boldsymbol{\pi}_k \mathbf{S}^{(j-k)} )
( \mathbf{S}^{(0)} )^{-1} ,
\end{equation*}
where $\mathbf{F}^{(0)} = \mathbf{L}$, $\hat{\mathbf{S}}^{(j)} = \textstyle\sum_{l=j}^{\infty} \hat{\mathbf{F}}^{(l)} \mathbf{G}^{l-j}$ for $j \geq1 $, and $\mathbf{S}^{(j)} = \textstyle\sum_{l=j}^{\infty} \mathbf{F}^{(l)} \mathbf{G}^{l-j}$ for $j \geq 0$.
Vector $\boldsymbol{\pi}_{0}$ is uniquely determined by the normalization condition, and it can be found by solving
\begin{equation*}
\boldsymbol{\pi}_0 [ ( \hat{\mathbf{L}} - \hat{\mathbf{S}}^{(1)}
( \mathbf{S}^{(0)} )^{-1}\mathbf{B} )^{\lozenge}
| \mathbf{1}^T - \mathbf{H1}^T ]
= [ \mathbf{0} | 1 ],
\end{equation*}
where $\mathbf{H} = \sum_{j=1}^{\infty} \hat{\mathbf{S}}^{(j)} (\sum_{j=0}^{\infty} \mathbf{S}^{(j)} )^{-1}$, $\hat{\mathbf{L}} = \hat{\mathbf{A}} - \mathbf{I}$, and the symbol $\lozenge$ denotes an operator that discards the last column of the corresponding matrix~\cite{Riska-sigmet02}.
\end{proposition}
\begin{IEEEproof}
A proof for an equivalent continuous-time formulation is available in~\cite{Riska-sigmet02}; it is based on solving $\boldsymbol{\pi} \tilde{\mathbf{T}} = \mathbf{0}$.
The discrete-time case can be obtained by defining $\tilde{\mathbf{T}} = \mathbf{T} - \mathbf{I}$, which leads to a solution for $\boldsymbol{\pi} \mathbf{T} = \boldsymbol{\pi}$, as desired.
\end{IEEEproof}

Matrix $\mathbf{G}$ admits a nice interpretation: entry $[\mathbf{G}]_{r,c}$ is the conditional probability that the Markov process first enters level $i-1$ through state $c$ given that it starts at level~$i$, in state~$r$~\cite{Riska-sigmet02}.
Such a matrix must naturally satisfy the relation
\begin{equation*}
\mathbf{A}\mathbf{G} + \mathbf{B}
+ \textstyle\sum_{j=1}^{\infty} \mathbf{F}^{(j)} \mathbf{G}^{j+1}
=\mathbf{G} ,
\end{equation*}
and this equation can be solved recursively, as described above.
Figure~\ref{fig:G} illustrates the probabilistic interpretation of $\mathbf{G}$ and its powers.
As a side note, we emphasize that all the matrix equations simplify to scalar computations for the binary symmetric channel.

\ifoneCol
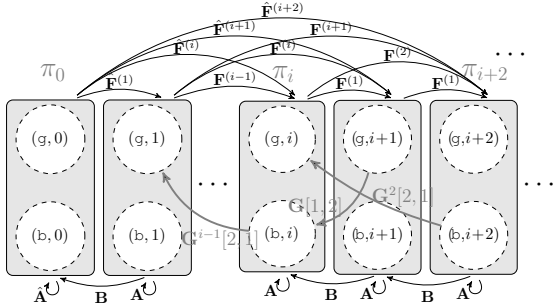
\begin{figure}
\centering
\scalebox{0.61}{\begin{tikzpicture}[->,>=stealth',shorten >=1pt,auto,node distance=0.6cm]
    \node (State00) [state2] {$\phantom{\!+\!}(\good,0)\phantom{1}$};
    \node (State01) [state2,below=of State00] {$\phantom{\!+\!}(\bad,0)\phantom{1}$} ;

    \begin{pgfonlayer}{background}

        \path (State00.west |- State00.north)+(-0.2,0.1) node (a) {};
        \path (State01.east |- State01.south)+(0.2,-0.1) node (b) {};
        \path (State00.east |- State00.north)+(-0.3,0.1) node (c) {};
        \path (State00.east |- State00.north)+(-0.7,0.7) node (po1) {\LARGE{$\color{gray}\mathbf{\pi}_0$}};\normalsize
        \path (State01.east |- State01.south)+(-0.7,-0.1) node (cc) {};
        \path [fill=gray!20,rounded corners, draw=black] (a) rectangle (b);

    \end{pgfonlayer}

      \node (State02) [state2,right=of State00] {$\phantom{\!+\!}(\good,1)\phantom{1}$};
      \node (State03) [state2,below=of State02] {$\phantom{\!+\!}(\bad,1)\phantom{1}$} ;

    \begin{pgfonlayer}{background}

        \path (State02.west |- State02.north)+(-0.2,0.1) node (d) {};
        \path (State03.east |- State03.south)+(0.2,-0.1) node (e) {};
        \path (State02.east |- State02.north)+(-0.3,0.1) node (f) {};
        \path (State02.east |- State02.north)+(-0.7,0.7) node (po2) {}; 
        \path (State03.east |- State03.south)+(-0.7,-0.1) node (ff) {};
        \path [fill=gray!20,rounded corners, draw=black] (d) rectangle (e);

    \end{pgfonlayer}

        \node (S2) [right=of State02] {};
        \node (I2)  [below=of S2]{\LARGE$\cdots$};\normalsize
        \node (u2) [below=of I2] {};
    \path (c) edge  [bend left=20] node[above=-3pt] {$\hat{\mathbf{F}}^{(1)}$} (f);
    \begin{pgfonlayer}{background}
        \path [fill=white,rounded corners, draw=white] (S2) rectangle (u2);
    \end{pgfonlayer}

        \node (State04) [state2,right=of S2] {$\phantom{\!+\!}(\good,i)\phantom{1}$};
        \node (State05) [state2,below=of State04] {$\phantom{\!+\!}(\bad,i)\phantom{1}$} ;

    \begin{pgfonlayer}{background}

        \path (State04.west |- State04.north)+(-0.2,0.1) node (g) {};
        \path (State05.east |- State05.south)+(0.2,-0.1) node (h) {};
        \path (State04.east |- State04.north)+(-0.3,0.1) node (i) {};
        \path (State04.east |- State04.north)+(-0.7,0.7) node (po3) {\LARGE{$\color{gray}\mathbf{\pi}_i$}};\normalsize
        \path (State05.east |- State05.south)+(-0.7,-0.1) node (ii) {};
        \path [fill=gray!20,rounded corners, draw=black] (g) rectangle (h);

    \end{pgfonlayer}
    \path (c) edge  [bend left=40] node[above=-3pt] {$\hat{\mathbf{F}}^{(i)}$} (i);
    \path (f) edge  [bend left=20] node[above=-3pt] {$\mathbf{F}^{(i-1)}$} (i);
    \path (ff) edge  [bend left=20] node[below] {$\mathbf{B}$} (cc);

        \node (State06) [state2,right=of State04] {$(\!\good,\!i\!+\!1)$};
        \node (State07) [state2,below=of State06] {$(\!\bad,\!i\!+\!1)$} ;

    \begin{pgfonlayer}{background}

        \path (State06.west |- State06.north)+(-0.2,0.1) node (j) {};
        \path (State07.east |- State07.south)+(0.2,-0.1) node (k) {};
        \path (State06.east |- State06.north)+(-0.3,0.1) node (l) {};
        \path (State07.east |- State07.south)+(-0.7,-0.1) node (ll) {};
        \path [fill=gray!20,rounded corners, draw=black] (j) rectangle (k);

    \end{pgfonlayer}
    \path (c) edge  [bend left=40] node[above=-3pt] {$\hat{\mathbf{F}}^{(i+1)}$} (l);
    \path (f) edge  [bend left=40] node[above=-3pt] {$\mathbf{F}^{(i)}$} (l);
    \path (i) edge  [bend left=20] node[above=-3pt] {$\mathbf{F}^{(1)}$} (l);
    \path (ll) edge  [bend left=20] node[below] {$\mathbf{B}$} (ii);

        \node (State08) [state2,right=of State06] {$(\!\good,\!i\!+\!2)$};
        \node (State09) [state2,below=of State08] {$(\!\bad,\!i\!+\!2)$} ;

    \begin{pgfonlayer}{background}

        \path (State08.west |- State08.north)+(-0.2,0.1) node (m) {};
        \path (State09.east |- State09.south)+(0.2,-0.1) node (n) {};
        \path (State08.east |- State08.north)+(-0.3,0.1) node (o) {};
        \path (State08.east |- State08.north)+(-0.5,0.7) node (po5) {\LARGE{$\color{gray}\mathbf{\pi}_{i+2}$}};\normalsize
        \path (State09.east |- State09.south)+(-0.7,-0.1) node (O) {};
        \path [fill=gray!20,rounded corners, draw=black] (m) rectangle (n);

    \end{pgfonlayer}
        \path (State09) edge  [bend left=10, very thick,draw=gray!100] node[near start=1pt,above] {\large$\color{gray!100}\mathbf{G}^2[2,1]$} (State04);\normalsize
        \path (State06) edge  [bend left, very thick,draw=gray!100] node[left] {\large$\color{gray!100}\mathbf{G}[1,2]$} (State05);\normalsize
        \path (State05) edge  [bend left, very thick,draw=gray!100] node[near start=2pt,below=-1pt] {\large$\color{gray!100}\mathbf{G}^{i-1}[2,1]$} (State02);\normalsize
    \path (c) edge  [bend left=40] node[above=-3pt] {$\hat{\mathbf{F}}^{(i+2)}$} (o);
    \path (f) edge  [bend left=40] node[above=-3pt] {$\mathbf{F}^{(i+1)}$} (o);
    \path (i) edge  [bend left=40] node[above=-3pt] {$\mathbf{F}^{(2)}$} (o);
    \path (l) edge  [bend left=20] node[above=-3pt] {$\mathbf{F}^{(1)}$} (o);
    \path (O) edge  [bend left=20] node[below] {$\mathbf{B}$} (ll);

    \path (ll) edge  [loop below,looseness=8,out=-60,in=-120] node[left] {$\mathbf{A}$} (ll);
    \path (O) edge   [loop below,looseness=8,out=-60,in=-120] node[left] {$\mathbf{A}$} (O);
    \path (ii) edge  [loop below,looseness=8,out=-60,in=-120] node[left] {$\mathbf{A}$} (ii);
    \path (ff) edge  [loop below,looseness=8,out=-60,in=-120] node[left] {$\mathbf{A}$} (ff);
    \path (cc) edge  [loop below,looseness=8,out=-60,in=-120] node[left] {$\hat{\mathbf{A}}$} (cc);
\node (S3) [right=of State08] {};
        \node (I3)  [below=of S3]{\LARGE$\cdots$};\normalsize
        \node (u3) [below=of I3] {};

    \begin{pgfonlayer}{background}
        \path [fill=white,rounded corners, draw=white, dashed] (S3) rectangle (u3);
    \end{pgfonlayer}
    \path (State08.east |- State08.north)+(0.1,1.1) node (o) {\LARGE$\cdots$};\normalsize
\end{tikzpicture}
}
\caption{\label{fig:G}Level transition diagram and probabilistic interpretation of $\mathbf{G}$}
\end{figure}
\fi

To conclude this section, we introduce a slight generalization of the arrival process.
Consider a two-state discrete-time Markov-modulated Poisson process with arrival rates $\lambda_1$ and $\lambda_2$, MMPP$(\lambda_1,\lambda_2)$.
The only elements of our analysis that need to be modified are the blocks in the transition probability matrix $\mathbf{T}$;
they become $4 \times 4$ matrices to account for the state of the modulating process.
Proposition~\ref{proposition:MMPP} offers a formal description of the quantities involved in making changes.

\begin{proposition} \label{proposition:MMPP}
Suppose that $T_1$ represents the amount of time the arrival process, MMPP$(\lambda_1, \lambda_2)$, spends in modulating state one during the transmission of a codeword.
The joint probability that $i$ packets arrive during that time interval together with the modulating process transitioning to state $A_{N+1}$, conditioned on starting state $A_1$, is
\begin{equation*}
P_{K_a,A_{N+1}\mid A_1}(i, l | m)
= \textstyle\sum_{t=0}^{N} P_{K_a|T_1}(i | t) P_{T_1, A_{N+1} | A_1}(t, l | m),
\end{equation*}
$l,m \in \{ 1, 2 \}$, where $K_a$ denotes the number of arrivals and $P_{T_1, A_{N+1} | A_1}(t, l | m)$ accounts for the occupation time of the modulating process as well as edge transitions~\cite[Lemma~1]{HamidiSepehr-aller12}.
The conditional distribution of arrivals, $P_{K_a|T_1}(i | t)$ becomes
\begin{equation*}
\sum_{k_1=0}^i \sum_{k_2=0}^{i-k_1}
\frac{(\lambda_1 t)^{k_1}}{k_1!} e^{-\lambda_1 t}
\frac{(\lambda_2 (N-t))^{k_2}}{k_2!} e^{-\lambda_2 (N-t)} .
\end{equation*}
\end{proposition}

Collecting these results, we gather that $a_i$ must be replaced by $P_{K_a,A_{N+1} | A_1}(i,l | m)$ in the transition probabilities of the queue, \eqref{eq:transProb}.
This yields $4\times4$ blocks in the modified transition probability matrix.
In the revised formulation,
\begin{equation*}
\boldsymbol{\pi}_q
= [\pi(4q+1) \; \pi(4q+2) \; \pi(4q+3) \; \pi(4q+4)] ,
\end{equation*}
which corresponds to having $q$ packets with a specific pair of channel state and modulating state for the arrival process.

\section{Probability of Decoding Failure}
\label{section:fail}

In this section, we derive probabilities of decoding failures for various scenarios.
We begin with the simpler BSC case, and then we proceed to the Gilbert-Elliott channel.

\subsection{Random Coding with ML/MD Decoding}

Consider a coding scheme in which a codebook of size $M=2^{NR}$ is generated at random.
As before, $R$ denotes code rate and $N$ stands for block length.
For every index $i \in \{ 1, \ldots, M \}$, a codeword $\mathbf{X}(i)$
is selected uniformly and independently from the set of length-$N$ binary sequences, $\{ 0, 1 \}^N$.
The maximum number of information bits encoded in each transmission is $K=\log_2 M$.
For performance assessment, we assume that one of the codewords is chosen at random and sent over the communication channel.
On the receiver side, a maximum likelihood (ML) decision rule is used to decode the received vector $\mathbf{Y}$; that is, $\hat{\mathbf{X}} = \operatorname{arg\,max}_{\mathbf{X}} P_{\mathbf{Y}|\mathbf{X}}(\mathbf{y}|\mathbf{x})$.

\subsubsection{Binary Symmetric Channel}
\label{sec:Binary-Symmetric-Channel}

For our memoryless channel, the ML decoder actually decodes to the closest valid codeword.
The ML decision rule is therefore equivalent to the minimum distance (MD) decoder.
A subtle, yet important point in analyzing this decoder is that the decoding radius is not fixed in advance; this can be inferred from the following well-known result.

\begin{thm}[\cite{Fano-1961}] \label{thm:Error-probability}
Random coding is employed to send information over a BSC$(p)$.
Assume that ML decoding is performed at the receiver, with ties treated as decoding failures.
Then, the failure probability for this scenario is given by
\begin{align}
\label{eq:main_sum}
&P_{\mathrm{f}} = \sum_{e=0}^{N} P_E(e) P_{\mathrm{f}|E}(e) \\
&= \sum_{e=0}^{N} \binom{N}{e}p^e(1-p)^{N\!-\!e}
\left[ 1 \! - \!
\left( \! 1 \! - \! 2^{-N} \sum_{i=0}^{e} \binom{N}{i} \right)^{M-1} \right] .
\nonumber
\end{align}
\end{thm}

We note this result holds for any forward error correction scheme in which all codewords are equally likely, and that they are pairwise independent (e.g., Shannon random coding or random linear codes).
Moreover, the format of \eqref{eq:main_sum} extends to other encoding strategies.
For a BSC$(p)$, random variable $E$ possesses a binomial distribution and a suitable expression for the conditional probabilities of decoding failure should be substituted.
In~\cite{Polyanskiy-it10}, Fano's result is modified to better handle ties, and it is generalized to a wider class of channels.
It turns out that, for our purpose, this modification has a negligible effect on performance; and it is therefore disregarded.

\subsubsection{Gilbert-Elliott Channel}

Having gained valuable insight with the BSC$(p)$, we turn to the more challenging case.
We derive probabilities of decoding failure for the Gilbert-Elliott channel under ML decoding, and conditioned on the occupancy times.
We emphasize that knowing the empirical channel state distribution is key in finding useful expressions for failure probabilities.
Let $N_{\good}$ and $N_{\bad} = N - N_{\good}$ represent the numbers of visits to each channel state during the transmission of a length-$N$ codeword.
These random variables are sometimes collectively called the channel state type~\cite{Cover-1991}.
Using the empirical state distribution and the corresponding conditional error probabilities, one can average over all channel types to get the probabilities of decoding failure while accounting for boundary states,
\begin{equation} \label{eq:1}
\!\!P_{\mathrm{f}, C_{N+1}|C_1}(d|c)
\!= \!\textstyle\sum_{n_\good=0}^N \!P_{\mathrm{f}|N_{\good}}(n_{\good})
P_{N_\good,C_{N+1}|C_1}(n_\good,d|c) ,\!\!\!\!
\end{equation}
where $P_{N_\good,C_{N+1}|C_1}(\cdot,\cdot|\cdot)$ is given by~\cite[Lemma~1]{HamidiSepehr-aller12}.
One can also compute this latter quantity using the $N$-th power of the matrix generating function of the good state occupation time.
\begin{equation*}
\renewcommand{\arraystretch}{0.5}
\mathbf{G}(x) = \begin{bmatrix} (1-\alpha)x & \alpha x \\
\beta  & 1-\beta \end{bmatrix} .
\end{equation*}
We stress that the failure probabilities depend on the initial and final states of the channel through the distribution of $N_{\good}$.
Since we are interested in moderate block lengths, on the order of the mixing time of the channel, these boundary states can have a significant impact on the probabilities of decoding failure.

For a specific channel realization, let $\mathbf{X}_c$ and $\mathbf{Y}_c$ be the subvectors of $\mathbf{X}$ and $\mathbf{Y}$ corresponding to time instants when the channel is in state $c \in \{\good,\bad\}$.
We denote the number of errors in state~$c$ by $E_c=d_{\mathrm{H}} (\mathrm{X}_c,\mathrm{Y}_c)$, where $d_{\mathrm{H}}(\cdot,\cdot)$ represents the Hamming distance.
Conditional error probability $P_{\mathrm{f}|N_{\good}}(n_{\good})$ can then be written as
\begin{equation} \label{eq:2}
\sum_{e_{\good}=0}^{n_{\good}} \sum_{e_{\bad}=0}^{n_{\bad}}
P_{\mathrm{f}|N_{\good}, E_{\good}, E_{\bad}}(n_{\good},e_{\good},e_{\bad})
P_{E_{\good},E_{\bad}|N_{\good}}(e_{\good},e_{\bad}|N_{\good}) .
\end{equation}
Given the channel type, the numbers of errors in the good and bad states are independent,
\begin{equation} \label{eq:3-1}
P_{E_{\good},E_{\bad}|N_\good}(e_\good,e_\bad|n_\good)
= P_{E_{\good}|N_\good}(e_\good|n_\good)P_{E_{\bad}|N_\good}(e_\bad|n_\good) ,
\end{equation}
where individual distributions are simply given by
\begin{equation} \label{eq:3}
P_{E_{c}|N_c}(e_c|n_c)
= \textstyle\binom{n_{c}}{e_{c}} \varepsilon_c^{e_{c}}(1-\varepsilon_{c})^{n_{c}-e_{c}}
\qquad c \in \{ \good, \bad\} .
\end{equation}

\begin{thm} \label{GEC_ML_BER}
When ties are treated as errors, the probability of decoding failure for a length-$N$ uniform random code with $M$ codewords, conditioned on the number of symbol errors in each state and the channel state type, is given by
\begin{equation} \label{eq:7}
P_{\mathrm{f}|N_{\good}, E_{\good}, E_{\bad}}(n_\good,e_\good,e_\bad)
\!= \!1-(1-\textstyle2^{-N}\!\textstyle\sum_{\mathcal{M}(\gamma e_{\good}+e_{\bad})}
\!\textstyle\binom{n_{\good}}{\tilde{e}_{\good}}\!\textstyle\binom{n_{\bad}}{\tilde{e}_{\bad}}\!)^{M-1} \!.
\end{equation}
where $\mathcal{M}(d)$ is the set of pairs $(\tilde{e}_{\good},\tilde{e}_{\bad})\in \{0,\ldots,N\}^2$ that satisfy $\gamma \tilde{e}_{\good} + \tilde{e}_{\bad} \leq d$.
This holds with $\gamma = \frac{\ln \varepsilon_{\good}-\ln(1-\varepsilon_{\good})}{\ln\varepsilon_{\bad}-\ln(1-\varepsilon_{\bad})}$ for the ML decision rule, and with $\gamma=1$ for the MD decoder.
\end{thm}
\begin{IEEEproof}
First, we revisit the ML decoding rule for the Gilbert-Elliott channel when channel state information is available at the receiver.
Given the state occupation $n_{\good}$, we have
\begin{align*}
P_{\mathbf{Y}|\mathbf{X}}(\mathbf{y}|\mathbf{x})
&= P_{\mathbf{Y}_{\good}|\mathbf{X}_{\good}}(\mathbf{y}_{\good}|\mathbf{x}_{\good})
P_{\mathbf{Y}_{\bad}|\mathbf{X}_{\bad}}(\mathbf{y}_{\bad}|\mathbf{x}_{\bad})\\
&=\varepsilon_{\good}^{e_{\good}}(1-\varepsilon_{\good})^{n_{\good}-e_{\good}}
\varepsilon_{\bad}^{e_{\bad}}(1-\varepsilon_{\bad})^{n_{\bad}-e_{\bad}}
\end{align*}
Upon receiving word $\mathbf{Y}$, the ML decoder returns the codeword $\mathbf{X}$ that maximizes $\ln P_{\mathbf{Y}|\mathbf{X}}(\mathbf{y}|\mathbf{x})$.
Thus, a little algebra shows the decoded message will be
\begin{equation} \label{eq:ML}
\operatorname*{arg\,min}_{\mathbf{x} \in \mathcal{C}}
[\gamma e_{\good}(\mathbf{x}) + e_{\bad}(\mathbf{x})],
\end{equation}
where $e_{\good}(\mathbf{x}) = d_{\mathrm{H}} (\mathbf{x}_{\good},\mathbf{y}_{\good})$ and $e_{\bad}(\mathbf{x}) = d_{\mathrm{H}} (\mathbf{x}_{\bad},\mathbf{y}_{\bad})$ are realizations of $E_{\good}$ and $E_{\bad}$, respectively.
This argument is used to demonstrate the dependency on $\mathbf{x}$.
Notice that the term $n_{\good} \ln (1-\varepsilon_{\good}) + n_{\bad} \ln (1-\varepsilon_{\bad})$ in $\ln P_{\mathbf{Y}|\mathbf{X}}(\mathbf{y}|\mathbf{x})$ does not change the ML decision.

Next, we consider the probability of failure for the decoding rule given in \eqref{eq:ML} when random codes are used.
In our system, decoding succeeds if and only if the correct codeword is returned as the unique minimizer in \eqref{eq:ML}.
The failure probability found in \eqref{eq:7} can be obtained in a few steps.
By symmetry, we can assume that the transmitted codeword $\mathbf{x}$ is the all-zero codeword.
The other $M-1$ codewords are drawn independently and uniformly.
For any received vector $\mathbf{y}$ that satisfies $E_{\good} = e_{\good}$ and $E_{\bad} = e_{\bad}$, decoding succeeds when every other codeword produces a strictly larger value for the cost function in~\eqref{eq:ML}.
A straightforward combinatorial argument shows that the number of codewords that meet this requirement is
\begin{equation} \label{eq:Vol}
V(n_{\good},n_{\bad},e_{\good},e_{\bad})
= \textstyle\sum_{(\tilde{e}_{\good},\tilde{e}_{\bad}) \in \mathcal{M}(\gamma e_{\good}+e_{\bad})}
\binom{n_{\good}}{\tilde{e}_{\good}} \binom{n_{\bad}}{\tilde{e}_{\bad}} .
\end{equation}
The probability that a uniformly chosen random vector falls in this set is $q=V(n_{\good},n_{\bad},e_{\good},e_{\bad})/2^N$.
Since codewords are independent, the failure probability is equal to $1\!-(1\!-q)^{M\!-1}$.

One can infer from \eqref{eq:ML} that, for the ML decision rule, errors in the bad state do not affect performance as much as errors in the good state.
This is because the decoder gives more weight to symbols that are received while the channel is in its good state, as they are deemed more reliable.
The MD decoder, on the other hand, only considers the total number of errors within a block, irrespective of the state they occur in.
That is, errors in either state cost the same and $\gamma=1$.
The terms over which the sum is taken need to be modified accordingly.
\end{IEEEproof}

In view of Theorem~\ref{GEC_ML_BER}, one can substitute the appropriate expressions for decoding performance into \eqref{eq:2} to get overall probabilities of decoding failure.

As a side note, Vandermonde's convolution identity implies that
\begin{equation*}
\textstyle\sum_{\tilde{e}_{\good}=0}^{e_{\good}+e_{\bad}}
\textstyle\sum_{\tilde{e}_{\bad}=0}^{e_{\good}+e_{\bad}-\tilde{e}_{\good}}
\textstyle\binom{n_{\good}}{\tilde{e}_{\good}} \binom{n_{\bad}}{\tilde{e}_{\bad}}
= \textstyle\sum_{j=0}^{e_{\good}+e_{\bad}} \binom{N}{j},
\end{equation*}
and therefore the volume expression in \eqref{eq:Vol} for MD decoding ($\gamma=1$) reduces to the volume computation associated with the binary symmetric channel.
Finally, we note that
\begin{equation*}
\textstyle\sum_{\mathcal{M}(\gamma e_{\good}+e_{\bad})}
\textstyle\binom{n_{\good}}{\tilde{e}_{\good}} \binom{n_{\bad}}{\tilde{e}_{\bad}}
+ \textstyle\sum_{\mathcal{\bar{M}}(\gamma e_{\good}+e_{\bad})}
\textstyle\binom{n_{\good}}{\tilde{e}_{\good}} \binom{n_{\bad}}{\tilde{e}_{\bad}}
=\textstyle 2^{N} ,
\end{equation*}
where $\bar{\mathcal{M}}(c)$ is the set of pairs $(\tilde{e}_{\good},\tilde{e}_{\bad}) \in \{0,\ldots,N\}^2$ that satisfy $\gamma \tilde{e}_{\good} + \tilde{e}_{\bad} > c$.

\subsection{BCH Coding with Bounded Distance Decoding}
\label{subsection:BCHstandard}

In this section, we present a more pragmatic facet of our inquiry.
We consider a primitive binary BCH code of minimum distance $d_{\min}$, which is capable of correcting up to $t=\left\lfloor \frac{d_{\min}-1}{2}\right\rfloor $ errors.
This entails having $N = 2^m - 1$, with $m \geq 2$, and a single optimal $K$ for each $d_{\min}$~\cite[p. 486]{Wicker-1995}.
We analyze the queueing behavior of the system in terms of the block length $N$ and the code rate $R={K}/{N}$.
The goal is to characterize the performance over admissible parameters.
At the receiver, the bounded distance decoder either declares a decoding success, or it detects a failure and requests a retransmission.
It is important to emphasize that, when the number of errors is greater than $t$, the decoder may be subject to an undetected error.
We discuss this issue in greater depth in Section~\ref{section:ue}.

For the binary symmetric channel, the conditional probability of failure in \eqref{eq:main_sum} is equal to $P_{\mathrm{f}|E}(e)=\mathds{1}_{\{ z \in \mathbb{Z} | z > t \}}(e)$, where $\mathds{1}_A(\cdot)$ is the standard indicator function of the set $A$.
Similarly, for the Gilbert-Elliott case, the average failure probability is given by
\begin{align*}
P_{\mathrm{f}}
&= \textstyle\sum_{c,d \in \{ \good, \bad \}} P_{C_1}(c)
P_{\mathrm{f},C_{N+1}|C_1}(d|c)\\
&=\textstyle\sum_{c,d \in \{\good, \bad \}} P_{C_1}(c)
\textstyle\sum_{e=1}^N P_{E,C_{N+1}|C_1}(e,d|c) P_{\mathrm{f}|E}(e) ,
\end{align*}
where $P_{\mathrm{f}|E}(e)$ appears above.
The expected success probability can be computed in an analog fashion, albeit replacing $P_{\mathrm{f}|E}(e)$ by $1 - P_{\mathrm{f}|E}(e)$.
The average service rate can be expressed as $\mu_N = \rho_r P_{\mathrm{s}}$ packets per codeword transmission, thereby implicitly setting a bound for system stability.

\section{Undetected Errors}
\label{section:ue}

A serious issue with pragmatic communication systems is the presence of undetected decoding failures.
In the present setting, this occurs when the receiver uniquely decodes to the wrong codeword.
For delay-sensitive applications, this problem is especially important because recovery procedures can lead to undue delay.
To address this issue, we apply standard techniques that help control the probability of admitting erroneous codewords \cite{Forney-it68,Hof-it10}.
This, in turn, leads to slight modifications to the performance analysis presented above.
The probability of undetected failure is a system parameter that must be set during the design phase of the system.

\subsection{Random Coding with ML/MD Decoding}
Under our aforementioned scheme, information is sent over the channel and the decoder reports the codeword with the minimum (weighted) distance to the received vector, as seen in \eqref{eq:ML}.
To reduce the probability of undetected error, we revisit the technique established in \cite{Forney-it68} regarding the error exponents, and introduce a safety margin $\nu$~.
This scheme and its ramifications are easiest to explain for the binary symmetric channel.
Recall that, for this simpler channel model, the ML and MD decision rules coincide.
Suppose that $d_{\mathrm{H}} (\hat{\mathbf{x}}, \mathbf{y}) = \hat{e}$, where $\hat{\mathbf{x}}$ is the closest codeword to received vector $\mathbf{y}$.
The enhanced decoder only returns $\hat{\mathbf{x}}$ when the distance between $\mathbf{y}$ and the next closest codeword is greater than $\hat{e} + \nu$.
If another codeword is present within distance $\hat{e} + \nu$, then the receiver declares a decoding failure.

As before, let $e$ denote the distance between the sent message and the received vector.
The performance associated with this procedure can be characterized by considering balls of radii $e-\nu$, $e$, and $e+\nu$ centered around the received vector.
Notice that, by construction, the transmitted codeword always lies in the last two balls.
To analyze the system, consider the list of all codewords contained in the ball of radius $e+\nu$.
If there is exactly one codeword on this list, it must be the correct one and it is returned successfully by the decoder.
On the other hand, if there are more than one codeword on the list, then a decoding failure (detected or undetected) will occur.
One can write the probability of this event as
\begin{align}
\textstyle P_{\mathrm{f}|E}(e)=1-(1-2^{-N}\sum_{i=0}^{e+\nu}\binom{N}{i})^{M-1}.\label{eq:fail-undet}
\end{align}

A detected failure takes place when the decoder elects not to output a candidate codeword.
The problem is setup so that the correct codeword is always on the list.
As such, an undetected failure can only occur when there is at least one other candidate inside the ball of radius $e-\nu$.
Note that this condition is necessary, but not sufficient;
multiple incorrect candidates can be found in proximity of the received vector in such a way that a failure is reported.
If there are only two codewords in the ball of radius $e$ and one of them is inside the ball of radius $e-\nu$, then the decoder will necessarily return the incorrect one.
If there are more than two codewords with the ball of radius $e$, then detected and undetected failures can occur, although for well-designed systems such events are very rare.
Collecting these observations, we can derive an upper bound for the probability of undetected failure,
\begin{equation} \label{eq:undet}
\textstyle P_{\mathrm{ue}} \!<\! \sum_{e=0}^N \binom{N}{e} p^e (1\!-p)^{N\!-e}
[1\!-(1\!-\frac{\sum_{i=0}^{e-\nu-1} \binom{N}{i}}{2^N} )^{M\!-1} ] .
\end{equation}
It may be instructive to point out that ties between the closest codewords are always treated as detected failures.
Also, the probability of undetected failure decreases rapidly as $\nu$ gets larger.
Thus, by choosing an appropriate value for $\nu$, one can manage the level of undetected failures and hence make the decoding process more robust, at the expense of a higher overall probability of failure.
Lastly, since the probability of undetected failure is typically much smaller than the probability of detected failure, we can upper bound the latter by $P_{\mathrm{f}}$ with a negligible penalty.

Much of the intuition developed under the binary symmetric channel applies to the Gilbert-Elliott model, with one important distinction related to weighted distance.
Indeed, for this more elaborate finite-state channel,
the ML decoder picks the codeword that minimizes the weighted distance found in \eqref{eq:ML}, $\gamma e_{\good}(\mathbf{x})+e_{\bad}(\mathbf{x})$.
Suppose that $B$ is the minimum weighted distance between the received vector and a codeword, and let $C$
be the weighted distance associated with the transmitted codeword.
To deal with the probability of undetected failure, the decoder declares a failure if there is another codeword of weighted distance at most $B + \nu$.
Otherwise, the best candidate codeword is returned.

Similar to the BSC case, performance can be analyzed by considering three balls, with respect to weighted distance, of radii $C-\nu$, $C$, and $C+\nu$ centered around the received vector.
Again, the transmitted codeword always resides in the last two balls.
If there are multiple codewords on the list of codewords in the ball of radius $C+\nu$, then a decoding failure will occur.
This happens with probability
\ifoneCol
\begin{align*}
\textstyle P_{\mathrm{f}|N_{\good},E_{\good},E_{\bad}}(n_{\good},e_{\good},e_{\bad})= \textstyle 1 - ( 1 - 2^{-N} \sum_{\mathcal{M}(\gamma e_{\good}+e_{\bad}+\nu)}
\textstyle\binom{n_{\good}}{\tilde{e}_{\good}}
\textstyle\binom{n_{\bad}}{\tilde{e}_{\bad}} )^{M-1}.
\end{align*}
\else
\begin{align*}
\textstyle P_{\mathrm{f}|N_{\good},E_{\good},E_{\bad}}(&n_{\good},e_{\good},e_{\bad})\\
&= \textstyle 1 - ( 1 - 2^{-N} \sum_{\mathcal{M}(\gamma e_{\good}+e_{\bad}+\nu)}
\textstyle\binom{n_{\good}}{\tilde{e}_{\good}}
\textstyle\binom{n_{\bad}}{\tilde{e}_{\bad}} )^{M-1}.
\end{align*}
\fi
The joint probability of decoding failure and ending in state $C_{N+1}$, conditioned on starting state $C_1$, denoted $P_{\mathrm{f},C_{N+1}|C_1}(d|c)$, is upper bounded by
\ifoneCol
\begin{align} \label{eq:err}
\textstyle\bar{P}_{\mathrm{f},C_{N+1}|C_1}(d|c)
= \textstyle\sum_{n_{\good}=0}^N \!\sum_{e_{\good}=0}^{n_{\good}}
\!\textstyle\sum_{e_{\bad}=0}^{n_{\bad}}&\textstyle\binom{n_{\good}}{e_{\good}} \textstyle\!\binom{n_{\bad}}{e_{\bad}}
\varepsilon_{\good}^{e_{\good}}(1\!\!-\!\varepsilon_{\good})^{n_{\good}\!-e_{\good}}\varepsilon_{\bad}^{e_{\bad}}(1\!\!-\!\varepsilon_{\bad})^{n_{\bad}\!-e_{\bad}}\nonumber\\
&\times
\textstyle P_{\mathrm{f}|N_{\good},E_{\good},E_{\bad}}(n_{\good},e_{\good},e_{\bad})
P_{N_{\good},C_{N+1}|C_1}(n_{\good},d|c) .
\end{align}
In a similar fashion, the joint probability of undetected failure accounting for boundary states, $P_{\mathrm{ue},C_{N+1}|C_1}(d|c)$, is upper bounded by
\begin{align} \label{eq:undet}
\textstyle\bar{P}_{\mathrm{ue},C_{N+1}|C_1}(d|c)
= \textstyle\sum_{n_{\good}=0}^N \!\sum_{e_{\good}=0}^{n_{\good}}
\!\textstyle\sum_{e_{\bad}=0}^{n_{\bad}} &\textstyle\binom{n_{\good}}{e_{\good}} \textstyle\!\binom{n_{\bad}}{e_{\bad}}
\varepsilon_{\good}^{e_{\good}}(1\!\!-\!\varepsilon_{\good})^{n_{\good}\!-e_{\good}}\varepsilon_{\bad}^{e_{\bad}}(1\!\!-\!\varepsilon_{\bad})^{n_{\bad}\!-e_{\bad}}\nonumber\\
&\times
\textstyle P_{\mathrm{ue}|N_{\good},E_{\good},E_{\bad}}(n_{\good},e_{\good},e_{\bad})
P_{N_{\good},C_{N+1}|C_1}(n_{\good},d|c) ,
\end{align}
\else
\begin{align} \label{eq:err}
\textstyle\bar{P}_{\mathrm{f},C_{N+1}|C_1}(d|c)
&= \textstyle\sum_{n_{\good}=0}^N \!\sum_{e_{\good}=0}^{n_{\good}}
\!\textstyle\sum_{e_{\bad}=0}^{n_{\bad}} \nonumber\\
&\textstyle\binom{n_{\good}}{e_{\good}} \textstyle\!\binom{n_{\bad}}{e_{\bad}}
\varepsilon_{\good}^{e_{\good}}(1\!\!-\!\varepsilon_{\good})^{n_{\good}\!-e_{\good}}\varepsilon_{\bad}^{e_{\bad}}(1\!\!-\!\varepsilon_{\bad})^{n_{\bad}\!-e_{\bad}}\times\nonumber\\
\textstyle P_{\mathrm{f}|N_{\good},E_{\good},E_{\bad}}&(n_{\good},e_{\good},e_{\bad})
P_{N_{\good},C_{N+1}|C_1}(n_{\good},d|c) .
\end{align}
In a similar fashion, the joint probability of undetected failure accounting for boundary states, $P_{\mathrm{ue},C_{N+1}|C_1}(d|c)$, is upper bounded by
\begin{align} \label{eq:undet}
\textstyle\bar{P}_{\mathrm{ue},C_{N+1}|C_1}(d|c)
&= \textstyle\sum_{n_{\good}=0}^N \!\sum_{e_{\good}=0}^{n_{\good}}
\!\textstyle\sum_{e_{\bad}=0}^{n_{\bad}} \nonumber\\
&\textstyle\binom{n_{\good}}{e_{\good}} \textstyle\!\binom{n_{\bad}}{e_{\bad}}
\varepsilon_{\good}^{e_{\good}}(1\!\!-\!\varepsilon_{\good})^{n_{\good}\!-e_{\good}}\varepsilon_{\bad}^{e_{\bad}}(1\!\!-\!\varepsilon_{\bad})^{n_{\bad}\!-e_{\bad}}\times\nonumber\\
\textstyle P_{\mathrm{ue}|N_{\good},E_{\good},E_{\bad}}&(n_{\good},e_{\good},e_{\bad})
P_{N_{\good},C_{N+1}|C_1}(n_{\good},d|c) ,
\end{align}
\fi	
where $P_{\mathrm{ue}|N_{\good},E_{\good},E_{\bad}}(n_{\good},e_{\good},e_{\bad})$ is equal to \[\textstyle  1 - ( 1 - 2^{-N} \textstyle\sum_{\mathcal{M}(\gamma e_{\good}+e_{\bad}-\nu)} \textstyle\binom{n_{\good}}{\tilde{e}_{\good}} \binom{n_{\bad}}{\tilde{e}_{\bad}})^{M-1}.\]
As before, the probability of undetected decoding failure diminishes as $\nu$ increases.
Also, for most systems, the probability of detected failure is well approximated by the upper bound $P_{\mathrm{f},C_{N+1}|C_1}(d|c)$ because undetected failures are very unlikely.

\subsection{BCH Codes with Bounded Distance Decoding}

Our BCH codes are decoded using bounded distance decoding.
It is possible to devise a safety margin and thereby reduce the probability of undetected decoding failures in this setting as well.
In this case, an undetected error occurs when the received vector lies in the decoding region of an incorrect codeword.
Therefore, shrinking the decoding regions of admissible codewords can prevent undetected failures.
Let $\nu$ denote the size of the safety margin, and assume that the desired error-correcting capability of the code is $t-\nu$ errors, where $t$ is defined in Section~\ref{subsection:BCHstandard}.
Under this slight modification, the decoder can detect up to $t+\nu$ symbol errors.

We assume that a codeword is mapped to the channel using a uniform random interleaver and, as such, all error patterns consisting of $e$ errors are equally probable~\cite{Elliott-bell63}.
This introduces a symmetry in the problem that facilitates analysis.
Without loss of generality, one can assume that the zero codeword is transmitted to the destination.
For this situation, an undetected error occurs whenever the Hamming distance between the received word and a nonzero codeword is less than $t-\nu$.

We consider the performance of this scheme for the binary symmetric channel first.
In~\cite{Kim-com96}, the probability of undetected error for bounded distance decoding is computed.
Using the enhanced detecting radius $t+\nu$ (instead of $t$), we can write $P_{\mathrm{ue}}=\sum_{e=t+\nu+1}^{N}W(e)P_{E}(e)$,
where $W(e)$ denotes the conditional decoder failure probability defined as the ratio of the number of weight $e$ error patterns lying within distance $t-\nu$ from a codeword over the total number of weight $e$ words in the entire space.
This can be written as
\begin{equation} \label{eq:probmeasure}
W(e) = \frac{\sum_{j=0}^{t-\nu} \sum_{l=e-j}^{e+j} A_{l} \binom{N-l}{{(j+e-l)}/{2}}
\binom{l}{{(j-e+l)}/{2}}} {\binom{N}{e}} ,
\end{equation}
where $A_l$ denotes the number of weight $l$ codewords in a BCH code space, designed to correct up to $t=\left\lfloor \frac{(t-\nu)+(t+\nu)}{2}\right\rfloor $ errors where $(t-\nu)+(t+\nu)=d_{\min}-1$.
In other words, we use the weight distribution of a $t$ error-correcting BCH code in our decoder design; however, by using the lower $t-\nu$ error correcting capability and $t+\nu$ error detecting capability, we get better performance in terms of undetected errors.

Still, a main issue with this expression is that the weight distributions for most BCH codes are not known.
Furthermore, when an expression is known~\cite{Kasami-ccma69}, it may be too complicated to integrate into our analysis.
Nevertheless, one can approximate the weight distribution of a binary primitive BCH code of length $N=2^m-1$ and designed distance $d_{\min}=2t+1$, where $2t-1<2^{\left\lceil {m}/{2} \right\rceil}+1$, by a binomial-like distribution as~\cite{Kim-com96},
\begin{equation} \label{eq:al}
A_l = \begin{cases}
\renewcommand{\arraystretch}{0.7}
\begin{array}{@{}ll}
1, & l=0 \\
0, & 1 \leq l < d_{\min} \\
2^{-mt} \binom{N}{l}(1+E_{l}), & d_{\min} \leq l \leq \left\lfloor \frac{N}{2}\right\rfloor \\
A_{N-l}, & \left\lfloor \frac{N}{2}\right\rfloor \leq l\leq N
\end{array}
\end{cases}
\end{equation}
where $E_{l}$ is an error term in the approximation of the weight distribution of the BCH code by a binomial distribution.
It has been shown that for moderately large block lengths, $E_{l}$ is negligible.
Consequently, $W(e)$ is well approximated by $2^{-mt}\sum_{j=0}^{t-\nu} \binom{N}{j}$.
As a result, the probability of undetected error is approximately
\[
\textstyle P_{\mathrm{ue}}\approx 2^{-mt}\sum_{j=0}^{t-\nu}\binom{N}{j}\sum_{e=t+\nu+1}^{N}P_{E}(e)
.\]
This interpretation generalizes to the Gilbert-Elliott channel, and the conditional probability of undetected error is equal to
\begin{equation*}
\textstyle P_{\mathrm{ue},C_{N+1}|C_1}(d|c)
= \textstyle\sum_{e=t+\nu+1}^N W(e) P_{E,C_{N+1}|C_1}(e,d|c) ,
\end{equation*}
where {$c,d\in\{\good,\bad\}$} and $W(e)$ is unchanged from \eqref{eq:probmeasure}.
Similar to the BSC case, this function is well approximated by
\[
\textstyle  2^{-mt}\!\textstyle\sum_{j=0}^{t-\nu}\!\binom{N}{j}\!\sum_{e=t+\nu+1}^{N}\!P_{E,C_{N+1}|C_1}(e,d|c).\]
This result is supported through numerical simulations.

\section{Performance Evaluation}
\label{sec:Numerical-Results}

In this section, we evaluate our proposed methodology using traffic parameters based on a voice over IP (VoIP) application for an EVDO system, a 3G component of CDMA2000~\cite{Ahson-2008}.
This system offers an uplink sector capacity of 500 Kb/s with 16 active users per sector~\cite{Bhushan-commag06}.
For a VoIP system with more users and lower per-user rates, this is somewhat optimistic.
As such, for illustrative purposes, we choose a total uplink rate of 460~Kb/s per sector; this gives a rate of $R_{\mathrm{b}}=28.75$ Kb/s for each active user.

The enhanced variable rate codec (EVRC), used by CDMA2000 systems for low bit-rate speech, generates a voice packet every 20~ms.
EVRC features four distinct frame types corresponding to different bit-rates: full rate gives $171$ bits, $1/2$ rate gives $80$ bits, $1/4$ rate gives $40$ bits, and $1/8$ rate gives $16$ bits.
Hereafter, we adopt the rough estimates of the relative frequencies for the speech coder states published in~\cite{Ahson-2008}.
Moreover, as the header size for voice packets are usually very large relative to the voice payload, we assume that ROHC compression is employed to reduce overhead to four bytes.
Under these parameters, the average size of a voice packet becomes $1/{\rho}=\sum_i f_i (l_i + \mbox{overhead})=88.55$~bits, where $f_i$ is the relative frequency of state $i$ and $l_i$ denotes the frame size for the same state.
The number of header bits in every segment is set to $h = 2$.
Throughout the numerical evaluation, packets are assumed to arrive according to a Poisson process.
Since packets are generated every $20$~ms, we find that $\lambda=50$ packets per second and we receive an average of ${50}/{R_{\mathrm{b}}}$ packets/channel use.

The choice of a Poisson arrival process (or MMPP) allows us to make fair comparisons between codes with different block lengths.
In particular, the rate $\lambda$ in packets per channel use is fixed, and arrivals in the queue correspond to the number of packets produced by the source during the transmission time of one codeword.
The marginal distribution of the sampled process is also Poisson with arrival rate $\lambda N$, in packets per codeword.
Given this framework, a prime goal is to minimize the tail probability of the queue over possible values for parameters $N$ and $K$.

One drawback associated with our closed-form approach is that handling undetected block errors in a realistic manner (e.g., via late detection when the packet CRC fails) is not possible.
Therefore, to facilitate the analysis, we assume the presence of a genie that informs the receiver when an undetected block decoding error occurs.
Still, we require that the system maintain a probability of undetected error less than some threshold, and we disregard $(N,K)$ parameter pairs that violate this constraint.
Then, we evaluate the tail probability of the queue (the probability that the number of packets in the queue exceeds a prescribed threshold $\tau$) over all admissible values of $N$, $K$, and $\nu$ satisfying the undetected error probability constraint.
More precisely, we perform a two-stage procedure.
During the initial phase, the algorithm finds the smallest admissible integer $\nu$ corresponding to each pair $(N,K)$, subject to the prescribed upper bound on $P_{\mathrm{ue}}(N,K,\nu)$.
Once this is accomplished, the tail probability of the queue $\sum_{i=2\tau+1}^{\infty} \pi(i)$ is evaluated for different $(N,K)$ pairs using the optimum value of $\nu$ found in the previous step.
We emphasize that distribution $\{ \pi(i) \}$ is an implicit function of $\nu(N,K)$ and $P_{\mathrm{f}}(N,K,\nu)$.
To perform this procedure, we first evaluate the undetected error probabilities for different rates and $\nu=0$.
In many cases, $\nu=0$ satisfies the constraint.
For rates with high probabilities of undetected error, we increase $\nu$ progressively as to reduce the corresponding probabilities of undetected failure.
We stress that this necessarily increases the overall probability of decoding failure, as seen in \eqref{eq:fail-undet}--\eqref{eq:undet}.
Since we are interested in keeping the latter probability as small as possible, we raise $\nu$ until the undetected-error requirement is met and then stop.
The proper value of $\nu$ is generally very small, which makes the task fast and convenient.
Note that the initial phase of the procedure can be carried out offline beforehand, whereas the parameters of the coding scheme can be selected based on the current system conditions.
Values of $N$ and $K$ for which this procedure gives poor performance are ignored.

For illustrative purposes we present the curves corresponding to the tail probability of the queue versus the code rates, for various block lengths.
This effectively helps to understand how the choice of the code parameters significantly affects the queueing performance.
Furthermore, these curves reveal the existence of an optimal code rate associated to each clock length, and an optimal block length over all possible code lengths.
As such, one can fairly pick the $(N,K)$ pair which results in the best queueing performance.

While numerically evaluating our proposed methodology, we consider two cases: random coding with ML decoding over the BSC, and BCH coding with bounded distance decoding over Gilbert-Elliott channel.
The concise size of this survey is due, primarily, to space limitations.
Nonetheless, we believe that the insights offered by these two cases are applicable to other scenarios as well.

\subsection{Random Codes over the Binary Symmetric Channel}

Let the channel bit error rate be $p=0.1$, which yields a capacity of $C=0.531$ bits per channel use, and suppose that the constraint on $P_{\mathrm{ue}}(N,K,\nu)$ is $5 \times 10^{-5}$.
We know that increasing code rate $R$ for a fixed block length decreases redundancy and therefore reduces the error-correcting capability of the code.
Thus, the probability of decoding failure found in \eqref{eq:main_sum} becomes larger.
At the same time, changes in code rate affect $\rho_r$, the probability with which a codeword contains the last parcel of information of a packet.
As this rate varies, these two effects alter the transition probabilities and, hence, the stationary distribution of the Markov chain in opposite ways.

\begin{figure}
\centering
	\newlength\figureheight
	\newlength\figurewidth
	\setlength\figureheight{3.5cm}
	\setlength\figurewidth{10cm}
	\scalebox{\myScl}{
%
%
%
%
\begin{tikzpicture}[font=\footnotesize]

\begin{semilogyaxis}[%
view={0}{90},
width=\figurewidth,
height=\figureheight,
scale only axis,
xmin=0, xmax=200,
xmajorgrids,
ymin=0.001, ymax=1,
yminorticks=true,
ylabel={$\text{Tail probability Pr(}{\it\text{Q >10}}\text{)}$},
ymajorgrids,
yminorgrids,
legend columns=2,
legend style={at={(0.99,0.03)},anchor=south east,nodes=right,/tikz/column 2/.style={column sep=5pt}}]

\addplot [
color=black,
solid,
mark=*,
mark options={solid}
]
coordinates{
 (11.5,0.06873)(12,0.03268)(12.5,0.07935)(13,0.04845)(13.5,0.06759)(14,0.04778)(14.5,0.6202)(15,0.6453) 
};

\addlegendentry{N=50};

\addplot [
color=black,
solid,
mark=star,
mark options={solid}
]
coordinates{
 (16.1,0.03694)(16.8,0.0207)(17.5,0.02377)(18.2,0.01449)(18.9,0.00927)(19.6,0.009706)(20.3,0.01218)(21,0.009919)(21.7,0.03043)(22.4,0.05331)(23.1,0.06599) 
};

\addlegendentry{N=70};

\addplot [
color=black,
solid,
mark=triangle,
mark options={solid,,rotate=90}
]
coordinates{
 (26,0.01382)(27,0.009794)(28,0.006986)(29,0.005499)(30,0.004209)(31,0.004307)(32,0.003802)(33,0.006944)(34,0.007984)(35,0.04123)(36,0.07144) 
};

\addlegendentry{N=100};

\addplot [
color=black,
solid,
mark=triangle,
mark options={solid}
]
coordinates{
 (40.3,0.004224)(41.6,0.002784)(42.9,0.002842)(44.2,0.003898)(45.5,0.004295)(46.8,0.01028)(48.1,0.01619)(49.4,0.1583)(50.7,0.3459) 
};

\addlegendentry{N=130};
\addplot [
color=black,
solid,
mark=triangle,
mark options={solid,,rotate=180}
]
coordinates{
 (43.5,0.009717)(45,0.006328)(46.5,0.004265)(48,0.003686)(49.5,0.002619)(51,0.002667)(52.5,0.003501)(54,0.0033)(55.5,0.008554)(57,0.04089)(58.5,0.09212) 
};

\addlegendentry{N=150};

\addplot [
color=black,
solid,
mark=asterisk,
mark options={solid}
]
coordinates{
 (54,0.009378)(55.8,0.006709)(57.6,0.004901)(59.4,0.004335)(61.2,0.003296)(63,0.002851)(64.8,0.003039)(66.6,0.004695)(68.4,0.006569)(70.2,0.0249)(72,0.2264) 
};

\addlegendentry{N=180};
\addplot [
color=black,
solid,
mark=diamond,
mark options={solid}
]
coordinates{
 (58,0.01692)(60,0.01239)(62,0.00918)(64,0.006891)(66,0.005266)(68,0.004162)(70,0.003537)(72,0.003)(74,0.003337)(76,0.005369)(78,0.01533)(80,0.09513) 
};

\addlegendentry{N=200};
\addplot [
color=black,
solid,
mark=x,
mark options={solid}
]
coordinates{
 (68.2,0.01265)(70.4,0.00975)(72.6,0.006725)(74.8,0.005352)(77,0.004403)(79.2,0.003902)(81.4,0.004047)(83.6,0.005106)(85.8,0.01165)(88,0.05302)(90.2,0.1311) 
};

\addlegendentry{N=220};
\addplot [
color=black,
solid,
mark=square,
mark options={solid}
]
coordinates{
 (80,0.01482)(82.5,0.01078)(85,0.008805)(87.5,0.007295)(90,0.005666)(92.5,0.005203)(95,0.005093)(97.5,0.007277)(100,0.01544)(102.5,0.06947)(105,0.6376) 
};

\addlegendentry{N=250};
\addplot [
color=black,
solid,
mark=o,
mark options={solid}
]
coordinates{
 (89.6,0.02342)(92.4,0.01948)(95.2,0.01494)(98,0.01163)(100.8,0.009277)(103.6,0.007787)(106.4,0.00787)(109.2,0.009195)(112,0.01587)(114.8,0.04927)(117.6,0.3296) 
};

\addlegendentry{N=280};

\addplot [
color=black,
solid,
mark=+,
mark options={solid}
]
coordinates{
 (99.2,0.04156)(102.3,0.03268)(105.4,0.02594)(108.5,0.02077)(111.6,0.01569)(114.7,0.01314)(117.8,0.01157)(120.9,0.01148)(124,0.01957)(127.1,0.04787)(130.2,0.2339) 
};

\addlegendentry{N=310};

\end{semilogyaxis}
\node [below] at (0.5\figurewidth,-0.3) {\footnotesize Information Bits per Block (Segment Length), $K$} ;
\end{tikzpicture}}
\caption{Probabilities of buffer overflow for random codes over the BSC as functions of $K$, subject to constraint $P_{\mathrm{ue}} \leq 5 \times 10^{-5}$.}
\label{fig:The-probability-of}
\end{figure}
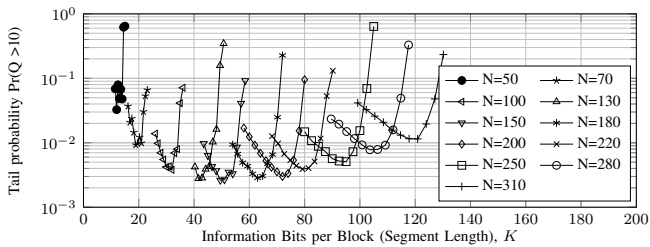

Figure~\ref{fig:The-probability-of} shows the complementary cumulative distribution functions evaluated at $\tau = 10$ packets as functions of $K$.
For each $(N,K)$ pair, $\nu$ has been chosen to satisfy the undetected error probability constraint, following the steps outlined above.
Each curve corresponds to a different block length and, as seen on the graph, there is a natural tradeoff between the probability of decoding failure and the payload per codeword.
For a fixed block length, neither the smallest segment length nor the largest one delivers optimal performance.
Moreover, block length must be selected carefully;
longer codewords do not necessarily yield better queueing performance.
For our system, optimal parameters are close to $(N,K) = (150,51)$, for which, the probability of undetected error is $3.67 \times 10^{-5}$, and $\nu=4$.

\subsection{BCH Codes over the Gilbert-Elliott Channel}

The parameters for our Gilbert-Elliott model are selected loosely based on QPSK modulation, a vehicular speed of $20$~mph, and a carrier frequency of $2.1$~GHz.
This gives a normalized Doppler frequency of $f_D T_s = 0.00082$, where $f_D$ represents the Doppler frequency and $T_s = {2}/{R_{\mathrm{b}}}$ is the symbol transmission time.
Setting the SNR threshold for transitions between the good and bad states to a common value of $\gamma_{\mathrm{th}}=2$~dB and the average received SNR to $\bar{\gamma}=15$~dB, we can apply the formulas given in \cite{Wilhelmsson-com99} and get model parameters
\begin{xalignat*}{2}
\alpha &= \textstyle\frac{\rho f_D T_s \sqrt{2\pi}}{e^{\rho^2}-1} = 0.3938 &
\beta = \textstyle\rho f_D T_s \sqrt{2\pi} = 0.0202
\end{xalignat*}
where $\rho=10^{{(\gamma_t - \bar{\gamma})}/{20}}$.
The probabilities of error in the good and bad states are chosen to be
\begin{align*}
\textstyle\varepsilon_{\good} &= \textstyle\frac{\alpha+\beta}{\alpha}
\textstyle\int_{\gamma_{\mathrm{th}}}^{\infty} f_{\Gamma}(\gamma) P_{\mathrm{e-QPSK}}(\gamma) d\gamma = 0.0097, \\
\textstyle\varepsilon_{\bad} &= \textstyle\frac{\alpha+\beta}{\beta}
\textstyle\int_0^{\gamma_{\mathrm{th}}} f_{\Gamma}(\gamma) P_{\mathrm{e-QPSK}}(\gamma) d\gamma = 0.3713,
\end{align*}
where $f_{\Gamma}(\cdot)$ is the probability distribution of the received SNR and $P_{\mathrm{e-QPSK}}(\gamma) = 1-(1-\mathcal{Q}(\sqrt{\gamma}))^2$ is the probability of symbol error for QPSK modulation.

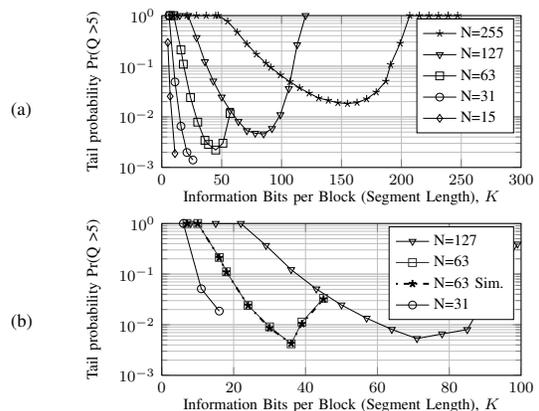
\begin{figure}
    \renewcommand{\arraystretch}{0.7}
	\centering
    \setlength\figureheight{2.75cm}
	\setlength\figurewidth{6.5cm}
    \begin{tabular}{m{0.5cm}m{7cm}}
        \scriptsize (a) &
        \scalebox{\myScl}{
%
%
%
%
\begin{tikzpicture}[font=\footnotesize]

\begin{semilogyaxis}[%
view={0}{90},
width=\figurewidth,
height=\figureheight,
scale only axis,
xmin=0, xmax=300,
xmajorgrids,
ymin=0.001, ymax=1,
yminorticks=true,
ylabel={$\text{Tail probability Pr(}{\it\text{Q >5}}\text{)}$},
ymajorgrids,
yminorgrids,
legend style={nodes=right,font=\footnotesize}]
\addplot [
color=black,
solid,
mark=star,
mark options={solid}
]
coordinates{
 (9,1)(13,1)(21,1)(29,1)(37,1)(45,1)(47,1)(55,0.76943646035756)(63,0.470001133425589)(71,0.275115661891873)(79,0.172566286417824)(87,0.11444141909011)(91,0.0948703219591129)(99,0.0672481850499263)(107,0.0493993698281477)(115,0.0374399156342056)(123,0.0293477789409159)(131,0.0235789688951265)(139,0.0211219298136079)(147,0.0189934735661838)(155,0.0182537402160122)(163,0.019154472468078)(171,0.0225547817937985)(179,0.0307757173552745)(187,0.0504096028954949)(191,0.107246181626936)(199,0.283413966316859)(207,1)(215,1)(223,1)(231,1)(239,1)(247,1) 
};

\addlegendentry{N=255};

\addplot [
color=black,
solid,
mark=triangle,
mark options={solid,,rotate=180}
]
coordinates{
 (8,1)(15,1)(22,1)(29,0.367253048348435)(36,0.122571264662709)(43,0.050859919322083)(50,0.0245403082814668)(57,0.0134588794820551)(64,0.00808059349814835)(71,0.00533121265468899)(78,0.00474386099388231)(85,0.00459927024318963)(92,0.00585952913353493)(99,0.0110092117535419)(106,0.0359258313418482)(113,0.263718253907707)(120,1) 
};

\addlegendentry{N=127};

\addplot [
color=black,
solid,
mark=square,
mark options={solid}
]
coordinates{
 (7,1)(10,1)(16,0.210494015041053)(18,0.109582944512851)(24,0.0240087131144102)(30,0.00785203725597615)(36,0.00343940144867735)(39,0.00285642726116676)(45,0.00221003596157235)(51,0.00301210681412015)(57,0.0115729545824303) 
};

\addlegendentry{N=63};

\addplot [
color=black,
solid,
mark=o,
mark options={solid}
]
coordinates{
 (6,1)(11,0.0489715195317797)(16,0.00653602727780399)(21,0.00197942944333646)(26,0.00140787207103853) 
};

\addlegendentry{N=31};

\addplot [
color=black,
solid,
mark=diamond,
mark options={solid}
]
coordinates{
 (5,0.293103785135775)(7,0.0254421401041771)(11,0.00188545749190446) 
};

\addlegendentry{N=15};

\end{semilogyaxis}
\node [below] at (0.5\figurewidth,-0.3) {\footnotesize Information Bits per Block (Segment Length), $K$} ;
\end{tikzpicture}}
        \tabularnewline \scriptsize (b) & \vspace{-0.1cm}
        \scalebox{\myScl}{
%
%
%
%
\begin{tikzpicture}[font=\footnotesize]

\definecolor{mycolor1}{rgb}{0.200000002980232,0.200000002980232,0.200000002980232}

\begin{semilogyaxis}[%
view={0}{90},
width=\figurewidth,
height=\figureheight,
scale only axis,
xmin=0, xmax=100,
xmajorgrids,
ymin=0.001, ymax=1,
yminorticks=true,
ylabel={\footnotesize{$\text{Tail probability Pr(}{\it\text{Q >5}}\text{)}$}},
ymajorgrids,
yminorgrids,
legend style={nodes=right,font=\footnotesize,/tikz/every even column/.append style={column sep=-0.3cm}}]

\addplot [
color=black,
solid,
line width=0.5pt,
mark=triangle,
mark options={solid,,rotate=180}
]
coordinates{
 (8,1)(15,1)(22,1)(29,0.367253048348435)(36,0.122571264662709)(43,0.050859919322083)(50,0.0245403082814668)(57,0.0134588794820551)(64,0.00808059349814835)(71,0.00533121265468899)(78,0.006596)(85,0.008006)(92,0.05773)(99,0.3923) 
};

\addlegendentry{N=127};
\addplot [
color=mycolor1,
solid,
line width=0.5pt,
mark=square,
mark options={solid}
]
coordinates{
 (7,1)(10,1)(16,0.2132)(18,0.1114)(24,0.02402)(30,0.0091)(36,0.0042)(39,0.0113)(45,0.0333) 
};

\addlegendentry{N=63};

\addplot [
color=black,
dash pattern=on 1pt off 3pt on 3pt off 3pt,
line width=1pt,
mark=star,
mark options={solid}
]
coordinates{
 (7,1)(10,1)(16,0.2105)(18,0.1096)(24,0.02401)(30,0.008672)(36,0.004248)(39,0.01045)(45,0.03204) 
};

\addlegendentry{N=63 Sim.};

\addplot [
color=black,
solid,
line width=0.5pt,
mark=o,
mark options={solid}
]
coordinates{
 (6,1)(11,0.05126)(16,0.0186) 
};

\addlegendentry{N=31};

\end{semilogyaxis}
\node [below] at (0.5\figurewidth,-0.3) {\footnotesize Information Bits per Block (Segment Length), $K$} ;
\end{tikzpicture}}
    \end{tabular}
\caption{\label{fig:Probability-of-Buffer} Probabilities of buffer overflow are displayed for various BCH codes over Gilbert-Elliott channel; (a) when undetected errors are not considered ($\nu=0$), (b)
when the decoding radius in every case is adjusted to meet the constraint on the probability of undetected error $P_{\mathrm{ue}} \leq 10^{-5}$.}
\end{figure}

This time, we require that the system features a probability of undetected error no greater than $10^{-5}$.
Recall that, for a specific $(N,K)$-BCH code, we can tradeoff the probability of misclassification and the ability to correct errors by changing the value of $\nu$.
Hence, we evaluate the tail probability of the queue over all admissible values of $N$, $K$, and $\nu$ satisfying the undetected error probability constraint.
To proceed, we first evaluate tail probabilities for admissible values of $N$ and $K$, with $\nu = 0$ (see Fig.~\ref{fig:Probability-of-Buffer}(a)).
Then, for the values of $K$ with high probabilities of undetected error, we increase $\nu$ progressively as to control misclassifications and meet the desired constraint.
Again, values of $N$ and $K$ that lead to inferior performance are discarded.
For example, for $N=63$, the values of $K=30,\,36,\,39,\,45$ are the ones with high probability of undetected error that are refined by increasing $\nu$ (see Fig.~\ref{fig:Probability-of-Buffer}(a)-(b)).
The values of $K$ greater than 45 associated to $N=63$ are ignored, since they result in poor performance after meeting the constraint on the  undetected errors.
Interestingly, for $N=63$, $K<30$, the constraint on the undetected errors is met with $\nu=0$.
Similar behavior is also observed for other block lengths.

The results associated with this procedure, in terms of the tail probability of the queue evaluated at $\tau=5$, are illustrated in Fig.~\ref{fig:Probability-of-Buffer}(b).
Comparing this graph to Fig.~\ref{fig:Probability-of-Buffer}(a), we gather that decreasing the likelihood of undetected error increases the tail probability of the queue.
In fact, because this forces the system to declare a detected error and request a retransmission more often, packets leave the queue less frequently.
Accordingly, the probability that the buffer exceeds a certain threshold goes up.
Looking at Fig.~\ref{fig:Probability-of-Buffer}(b), we see that the optimal code parameters are $(N,K)=(63,36)$.
The corresponding probability of undetected error is $8.78 \times 10^{-6}$ and $\nu=1$.
We note that the tail probability for $(N,K)=(127,71)$ is close to this optimal value.
This alternate configuration features an undetected error probability of $3.80 \times 10^{-8}$, which is achieved with $\nu=0$.
This survey demonstrates the need to adjust the value of $\nu$ on a per code basis.
Moreover, the results suggest that the proper value of $\nu$ is very small relative to $N$.

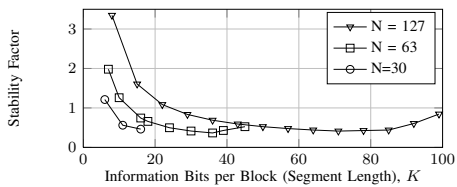
\begin{figure}[t]
\centering
  \setlength\figureheight{2.5cm}
  \setlength\figurewidth{6.5cm}
  \scalebox{\myScl}{
%
%
%
%
\begin{tikzpicture}[font=\footnotesize]

\begin{axis}[%
view={0}{90},
width=\figurewidth,
height=\figureheight,
scale only axis,
xmin=0, xmax=100,
xmajorgrids,
ymin=0, ymax=3.5,
ylabel={Stability Factor},
ymajorgrids,
legend style={nodes=right,font=\footnotesize}]
\addplot [
color=black,
solid,
line width=0.5pt,
mark=triangle,
mark options={solid,,rotate=180}
]
coordinates{
 (8,3.34676581038371)(15,1.60615076665577)(22,1.0849867356309)(29,0.834810458451475)(36,0.688426642797548)(43,0.592551213513294)(50,0.525234998248256)(57,0.476903786068639)(64,0.440368051878201)(71,0.413369283219224)(78,0.4269)(85,0.4397)(92,0.6051)(99,0.8445) 
};

\addlegendentry{N = 127};
\addplot [
color=black,
solid,
line width=0.5pt,
mark=square,
mark options={solid}
]
coordinates{
 (7,1.98109341735212)(10,1.25922927794366)(16,0.744102908692869)(18,0.658515633766033)(24,0.497965734828442)(30,0.4147)(36,0.3658)(39,0.4286)(45,0.5246) 
};

\addlegendentry{N = 63};

\addplot [
color=black,
solid,
line width=0.5pt,
mark=o,
mark options={solid}
]
coordinates{
 (6,1.2126)(11,0.5618)(16,0.4626) 
};

\addlegendentry{N=30};

\end{axis}
\node [below] at (0.5\figurewidth,-0.3) {\footnotesize Information Bits per Block (Segment Length), $K$} ;
\end{tikzpicture}}
\caption{This figure shows stability factors as functions of BCH code parameters;
when this factor exceeds one, the system is unstable.}
\label{fig:Updated-curves-for-1}
\end{figure}

Figure~\ref{fig:Updated-curves-for-1} plots the stability factor for the $(N,K)$ pairs found in our previous graph.
Systems for which ${\lambda_N}/{\mu_N}$ is larger than one are unstable.
We note that the tail probability is a good predictor of stability.
In general, systems with small stability factors feature good delay profiles as well.



Monte Carlo simulations provide additional empirical evidence for our proposed methodology.
This is especially important because our analysis assumes the existence of a genie that reports undetected errors.
To understand the effect of the genie, we perform simulations with and without the genie.
As expected, the genie-aided simulation results match our analysis almost perfectly.
In the absence of a genie, we assume that an undetected decoding error is eventually revealed by the packet CRC.
So long as the probabilities of undetected error remain relatively small, our simulations without the genie agree with both the coding and queueing performance predicted by the analytical framework.
For instance, Fig.~\ref{fig:Probability-of-Buffer}(b) superimposes simulation results for $N=63$ without the genie (dashed curve).
The plotted curves in this case are nearly indistinguishable.

\begin{figure}[tb]
\renewcommand{\arraystretch}{0.4}
\centering
\ifoneCol
	\setlength\figureheight{3.0cm}
	\setlength\figurewidth{5.5cm}
\else
	\setlength\figureheight{2.5cm}
	\setlength\figurewidth{4.9cm}
\fi	
\begin{tabular}{@{}l@{}@{}l@{}}
\scalebox{\myScl}{
%
%
%
%
\definecolor{mycolor1}{rgb}{1,1,0.862745098039216}
\begin{tikzpicture}[font=\footnotesize]

\begin{semilogyaxis}[%
width=\figurewidth,
height=\figureheight,
scale only axis,
xmin=0, xmax=13,
xmajorgrids,
ymin=3e-006, ymax=1,
yminorticks=true,
ymajorgrids,
yminorgrids,
legend style={draw=black,fill=white,align=left,font=\scriptsize}]
\addplot [
color=black,
solid,
line width=1.0pt,
mark=o,
mark options={solid}
]
coordinates{
 (0,1)(1,0.4164717)(2,0.1602558)(3,0.0612463)(4,0.0234831)(5,0.0089592)(6,0.0033896)(7,0.0012689)(8,0.0004697)(9,0.0001806)(10,6.98e-005)(11,2.33e-005)(12,8e-006)(13,2.7e-006) 
};
\addlegendentry{Geometric Packet Length};

\addplot [
color=black,
solid,
line width=1.0pt,
mark=triangle,
mark options={solid,,rotate=180}
]
coordinates{
 (0,1)(1,0.4518996)(2,0.1393166)(3,0.0365434)(4,0.0091467)(5,0.0022804)(6,0.0005633)(7,0.0001446)(8,3.99e-005)(9,1.2e-005)(10,3.6e-006)(11,1.8e-006) 
};
\addlegendentry{Constant Packet Length};
\addplot [
color=white,
mark size=3.5pt,
only marks,
mark=square*,
mark options={solid,fill=black,draw=mycolor1}
]
coordinates{
 (6,0.0005633) 
};
\end{semilogyaxis}
\draw [fill=white] (0.1cm,0.1cm) rectangle (1.5cm,0.7cm) ;
\node at (0.8cm,0.4cm) {\footnotesize $K=30$} ; 
\node [below] at (2.5cm,-0.2cm) {\footnotesize $\tau$ (packets)} ; 
\end{tikzpicture}%
}
&
\scalebox{\myScl}{
%
%
%
%
\definecolor{mycolor1}{rgb}{1,1,0.862745098039216}
\begin{tikzpicture}[font=\footnotesize]

\begin{semilogyaxis}[%
width=\figurewidth,
height=\figureheight,
scale only axis,
xmin=0, xmax=11,
xmajorgrids,
ymin=3.6e-006, ymax=1,
yminorticks=true,
ymajorgrids,
yminorgrids,
legend style={draw=black,fill=white,align=left,font=\scriptsize}]
\addplot [
color=black,
solid,
line width=1.0pt,
mark=o,
mark options={solid}
]
coordinates{
 (0,1)(1,0.3675497)(2,0.1221656)(3,0.0400144)(4,0.0130567)(5,0.0042457)(6,0.0014065)(7,0.0004684)(8,0.0001504)(9,5.21e-005)(10,1.92e-005)(11,6.2e-006)(12,3.3e-006)(13,4e-007)(14,0) 
};
\addlegendentry{Geometric Packet Length};

\addplot [
color=black,
solid,
line width=1.0pt,
mark=triangle,
mark options={solid,,rotate=180}
]
coordinates{
 (0,1)(1,0.3515886)(2,0.0797712)(3,0.0148273)(4,0.0025666)(5,0.0004273)(6,5.91e-005)(7,7.4e-006)(8,1.3e-006) 
};
\addlegendentry{Constant Packet Length};
\addplot [
color=white,
mark size=3.5pt,
only marks,
mark=square*,
mark options={solid,fill=black,draw=mycolor1}
]
coordinates{
 (6,5.91e-005) 
};
\end{semilogyaxis}
\draw [fill=white] (0.1cm,0.1cm) rectangle (1.5cm,0.7cm) ;
\node at (0.8cm,0.4cm) {\footnotesize $K=36$} ; 
\node [below] at (2.5cm,-0.2cm) {\footnotesize $\tau$ (packets)} ; 
\end{tikzpicture}%
}
\\
\scalebox{\myScl}{
%
%
%
%
\definecolor{mycolor1}{rgb}{1,1,0.862745098039216}
\begin{tikzpicture}[font=\footnotesize]

\begin{semilogyaxis}[%
width=\figurewidth,
height=\figureheight,
scale only axis,
xmin=0, xmax=12,
xmajorgrids,
ymin=3e-006, ymax=1,
yminorticks=true,
ymajorgrids,
yminorgrids,
legend style={draw=black,fill=white,align=left,font=\scriptsize}]
\addplot [
color=black,
solid,
line width=1.0pt,
mark=o,
mark options={solid}
]
coordinates{
 (0,1)(1,0.4306071)(2,0.1717854)(3,0.0680776)(4,0.0268821)(5,0.0105324)(6,0.0041439)(7,0.0016668)(8,0.000659)(9,0.0002622)(10,0.0001011)(11,3.79e-005)(12,1.38e-005)(13,3e-006) 
};
\addlegendentry{Geometric Packet Length};

\addplot [
color=black,
solid,
line width=1.0pt,
mark=triangle,
mark options={solid,,rotate=180}
]
coordinates{
 (0,1)(1,0.4399823)(2,0.137375)(3,0.0373407)(4,0.0096245)(5,0.0024131)(6,0.0006019)(7,0.0001504)(8,3.91e-005)(9,1.17e-005)(10,4e-006)(11,1.1e-006) 
};
\addlegendentry{Constant Packet Length};
\addplot [
color=white,
mark size=3.5pt,
only marks,
mark=square*,
mark options={solid,fill=black,draw=mycolor1}
]
coordinates{
 (6,0.0006019) 
};
\end{semilogyaxis}
\draw [fill=white] (0.1cm,0.1cm) rectangle (1.5cm,0.7cm) ;
\node at (0.8cm,0.4cm) {\footnotesize $K=39$} ; 
\node [below] at (2.5cm,-0.2cm) {\footnotesize $\tau$ (packets)} ; 
\end{tikzpicture}%
}
&
\scalebox{\myScl}{
%
%
%
%
\definecolor{mycolor1}{rgb}{1,1,0.862745098039216}
\begin{tikzpicture}[font=\footnotesize]

\begin{semilogyaxis}[%
width=\figurewidth,
height=\figureheight,
scale only axis,
xmin=0, xmax=15,
xmajorgrids,
ymin=1e-007, ymax=1,
yminorticks=true,
ymajorgrids,
yminorgrids,
legend style={draw=black,fill=white,align=left,font=\scriptsize},
legend pos= south west]
\addplot [
color=black,
solid,
line width=1.0pt,
mark=o,
mark options={solid}
]
coordinates{
 (0,0.999999999999996)(1,0.527432120583917)(2,0.263075833615211)(3,0.130974989718436)(4,0.065410052124254)(5,0.032430612526224)(6,0.016065287804255)(7,0.008034862686585)(8,0.004063095358916)(9,0.002007642066262)(10,0.001015487544959)(11,0.000529215880139)(12,0.000267685608833)(13,0.000147012363781)(14,8.4600104181e-005)(15,5.7831543298e-005)(16,4.0367562401e-005)(17,2.505079227e-005)(18,1.5746212283e-005)(19,8.15940091e-006)(20,3.435537225e-006) 
};
\addlegendentry{Geometric Packet Length};

\addplot [
color=black,
solid,
line width=1.0pt,
mark=triangle,
mark options={solid,,rotate=180}
]
coordinates{
 (0,1)(1,0.6084941)(2,0.2987297)(3,0.1362016)(4,0.0607781)(5,0.0270512)(6,0.0120838)(7,0.0054333)(8,0.002396)(9,0.0010306)(10,0.0004289)(11,0.0001773)(12,7.73e-005)(13,3.24e-005)(14,1.43e-005)(15,6.7e-006)(16,5.7e-006)(17,3.4e-006)(18,1.7e-006)(19,1e-006)(20,0) 
};
\addlegendentry{Constant Packet Length};
\addplot [
color=white,
mark size=3.5pt,
only marks,
mark=square*,
mark options={solid,fill=black,draw=mycolor1}
]
coordinates{
 (6,0.012084) 
};
\end{semilogyaxis}
\draw [fill=white] (3.4cm,1.8cm) rectangle (4.8cm,2.4cm) ;
\node at (4.2cm,2.1cm) {\footnotesize $K=45$} ; 
\node [below] at (2.5cm,-0.2cm) {\footnotesize $\tau$ (packets)} ; 
\end{tikzpicture}%
}
\end{tabular}
\caption{CCDF of stationary distribution of the queue length ($\Pr(\mbox{queue length}>\tau)$), is displayed for geometric and constant packet length, $N=63$. The tail probability of the queue for $\tau=5$ , has been marked with black squares in case of constant packet size.}\label{fig:ccdf}
\end{figure}
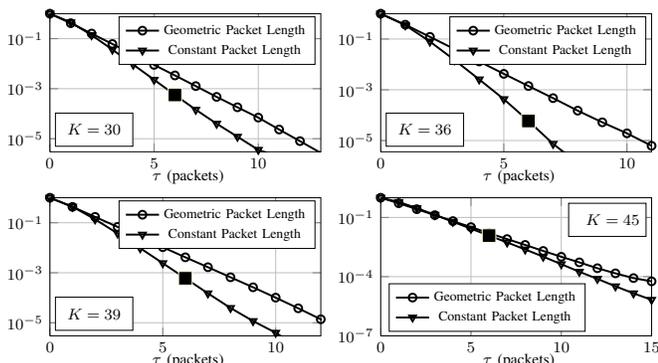

Another important concern pertains to possible modeling inaccuracies related to the traffic or the channel.
To examine such limitations, we carry Monte Carlo simulations for a system with constant packet lengths, $L=90$.
Figure~\ref{fig:ccdf} demonstrates the results in terms of the complementary cumulative distribution function (CCDF) of the queue occupancy for $N=63$ and different values of $K$.
We compare the results with those obtained for systems with geometric packet distributions, matching the means.
Not surprisingly, reducing variations in the arrival process decreases the tail probability of the queue.
That is, it makes the probability of a long queue very small.
This behavior should be expected since fixing the packet size precludes the arrival of a very long packet, an event that exacerbates the distribution of the queue.
In other words, designing the system using a geometric packet distribution leads to a conservative performance assessment compared to using a constant packet length.
Empirically, the system performs uniformly better in the latter case.
In a similar manner, smoothing the arrival process over time (e.g., periodic arrivals) should lead to a better profile.

Performance prediction aside, our analytical framework affords an efficient and accurate means of selecting system parameters.
For example, under stated channel conditions and queueing objectives, the optimum values for $N$ and $K$ are the same for constant and geometric packet length distributions.
Specifically, the minimum tail probability associated with the abstract model is achieved at $N=63$ and $K=36$.
Simulation results with constant packet sizes lead to the same operating point, although this latter approach is much more computationally demanding.
Altogether, simulation results offer strong support for the proposed methodology.

\section{Conclusions}
\label{section:Conclusion}

In this article, we introduce a novel framework to study the queueing behavior of coded wireless communications over finite-state error channels.
Through this framework, it is possible to select the optimal the block length and code rate of the encoding scheme based on the requirements of the system.
This is especially useful in the context of delay-sensitive applications for which long block lengths are inadequate.
The proposed methodology applies to both memoryless channels and channels with memory.
Due attention is given to undetected decoding failures, as they can have a very detrimental impact on the operation of pragmatic systems.
By using a safety margin, one can limit the likelihood of such events and thereby ensure adequate performance.

For illustrative purposes, a VoIP application is considered.
Channel parameters are derived from the CDMA2000 family of 3G mobile technology standards.
The proposed methodology enables the numerical evaluation of the equilibrium queue distribution.
This, in turn, can be employed to compute the tail probabilities of the queue occupancy and, subsequently, find the optimal operating point.
Our framework supports the rigorous comparison of coding schemes with different block lengths and code rates.
This study suggests that, for fixed conditions, optimal system parameters are essentially unaffected by small variations in the buffer overflow threshold.
The results and assumptions associated with our methodology are supported by Monte Carlo simulations.
This technique can be employed to facilitate adaptive modulation schemes that take into account both the channel profile and the requirements of the underlying traffic.
The optimization task can be carried out offline beforehand, whereas the parameters of the coding scheme can be selected based on current system conditions.
Possible avenues of future research include better accounting for feedback and extending this type of analysis to multi-user environments.

\vspace{-5pt}
\bibliographystyle{IEEEtr}
\bibliography{IEEEabrv,IEEEfull,WCLabrv,WCLfull,WCLbib,WCLnewbib}
\end{document}